\documentclass[aps,groupedaddress,preprint,superscriptaddress,showpacs]{revtex4-1}
\usepackage{eurosym}
\usepackage{amsmath}
\usepackage{fixmath}
\usepackage{graphicx}
\usepackage{wrapfig}
\usepackage[toc,page]{appendix}
\usepackage{subcaption}
\usepackage{hyperref}
\usepackage[capitalise]{cleveref}
\usepackage{subcaption}
\usepackage{amsfonts}
\usepackage{xcolor}

\graphicspath{{images/}}

\newcommand{\be}{\begin{equation}}
\newcommand{\ee}{\end{equation}}

\begin{document}

\title{Quantum dynamics of open many-qubit systems strongly coupled to a quantized electromagnetic field in dissipative cavities}

\author{ Mikhail Tokman}
\affiliation{Institute of Applied Physics, Russian Academy of Sciences, Nizhny Novgorod, 603950, Russia }
\author{Qianfan Chen}
\affiliation{Department of Physics and Astronomy, Texas A\&M University, College Station, TX, 77843 USA}
\author{ Maria Erukhimova}
\affiliation{Institute of Applied Physics, Russian Academy of Sciences, Nizhny Novgorod, 603950, Russia }
\author{Yongrui Wang}
\affiliation{Department of Physics and Astronomy, Texas A\&M University, College Station, TX, 77843 USA}
\author{Alexey Belyanin}
\affiliation{Department of Physics and Astronomy, Texas A\&M University, College Station, TX, 77843 USA}

\begin{abstract}

We study quantum dynamics of many-qubit systems strongly coupled to a quantized electromagnetic cavity mode, in the presence of decoherence and dissipation for both fermions and cavity photons.  The analytic solutions are derived for a broad class of open quantum systems in Lindblad approximation. They include identical qubits, an ensemble of qubits with a broad distribution of transition frequencies, and multi-level electron systems. Compact analytic solutions for time-dependent quantum state amplitudes and observables become possible with the use  of the stochastic equation  of evolution for the state vector. We show that depending on the initial quantum state preparation, the systems can evolve into a rich variety of entangled states with destructive or constructive interference between the qubits. In particular, dissipation in a cavity can drive the system into the dark states completely decoupled from the cavity modes. We also find the regimes in which multi-electron systems with a broad distribution of transition frequencies couple to the quantized cavity field as a giant collective dipole.

\end{abstract}

\date{\today }

\maketitle

\section{Introduction}

Solid-state cavity quantum electrodynamics (QED) attracted much interest as a promising platform
for quantum information and quantum sensing systems; see, e.g., \cite{thorma2015, lodahl2015, degen2017,  dovzhenko2018, bitton2019} for recent reviews. A typical scenario
involves an ensemble of quantum emitters (ideally, two-level systems) such
as quantum dots, defects in crystals, or molecules, strongly coupled to a
quantized electromagnetic (EM) field in a dielectric or plasmonic cavity. We will call these quantum
emitters qubits for brevity, to be distinguished from the logical qubits which may involve many
two-level systems as well as photonic or mixed degrees of freedom. Several
or many qubits are required for most applications. Although direct
near-field coupling of qubits with each other is possible and desired for some gating protocols, in the nanophotonics context 
such a coupling would require deterministic placement of qubits with sub-nm
accuracy, which is challenging. A simpler scenario which still permits various ways of the quantum state manipulation is the one in which the
qubits are coupled only through the common cavity mode(s). This is precisely the situation considered in this paper. We  derive step by step analytic solutions for the quantum dynamics of an ensemble of qubits strongly coupled to a quantized cavity mode, in the presence of decoherence and coupling to dissipative reservoirs, starting from the simplest case of $N$ identical qubits, then including a spread of transition frequencies in two or more two-level qubits, and finally the multilevel fermionic systems with arbitrary energy level distribution in the bands. Compact analytic solutions for the quantum states and observables in strongly coupled open quantum systems become possible within the formalism of the stochastic equation for the state vector which we recently developed \cite{tokman2020, chen2021}. 

As we show in the paper, this approach brings significant advantages as compared to the existing methods of treating open quantum systems. Indeed, the Heisenberg-Langevin formalism faces challenges when dealing with nonperturbative dynamics of strongly coupled systems as it leads to nonlinear operator-valued equations. The master equation formalism quickly becomes intractable analytically and allows only numerical treatment as the number of the degrees of freedom increases; see Appendix A for a detailed comparison.  Furthermore, determining the measure of entanglement within the master equation formalism is a separate problem since convenient universal figures of merit exist only for simplest systems. In contrast, our approach gives an analytic solution for a nonperturbative evolution of a strongly coupled system with an arbitrary number of the degrees of freedom. Furthermore, the explicit expression for the state vector makes the determination of entanglement or lack thereof much more transparent; it allows one to compare the resulting quantum state with Bell states and other quantum states popular in quantum information applications; it shows how the observables characterizing a given state gets affected by relaxation and eventually drown in an increasing noise, etc. Unlike the Wigner-Weisskopf  method, our approach treats the effects of noise correctly and yields a correct equilibrium state at long times.

Of course the possibility of analytic treatment comes with its limitations. We use the rotating wave approximation (RWA) throughout the paper, which means that all frequency scales such as Rabi frequencies and detunings are much smaller than the cavity mode frequency and the optical transition frequencies in the qubits. Beyond RWA, general analytic solutions become impossible in the fully quantized and nonlinear (nonperturbative) regime of light-matter interaction between fermionic and bosonic fields. Note that some recent experiments with low-frequency transitions (e.g. microwave or terahertz) in strongly coupled multi-electron systems went beyond the RWA into the ultra strong coupling regime; e.g., \cite{todorov2010, forndiaz2017, kono2019}.  

The simplest situation is when the qubits are identical, for example if the
qubit is a particular transition between two electron or vibrational states
in a given molecule and the system includes several or many identical molecules. We solve it in Sec.~II in the dissipationless case and then in Secs.~IV and V in the presence of dissipation and decoherence. It is
common knowledge \cite{thorma2015, dovzhenko2018, haroche} that  a system of $N$ identical qubits responds to the EM field as one qubit with a ``giant'' collective dipole moment which scales as $\sqrt{N}
$. At the same time, the qubits themselves demonstrate a much more complex
dynamics. Depending on the initial excited quantum state of the qubits,  they may develop either constructive or destructive interference during their evolution, which leads to either a complete energy transfer to the cavity mode or a complete suppression of such transfer.  For example, a system in which only one qubit is initially
excited may evolve into an entangled state of all qubits which is decoupled
from the cavity field. Moreover, in the typical case relevant for the plasmonic nanocavity \cite{chikkaraddy2016,benz2016, park2016, pelton2018, gross2018, park2019},  when the dissipation of
a cavity mode  is much faster than the decoherence
in a qubit, the field dissipation inevitably drives the system
into such a decoupled entangled state which remains preserved until the
 dissipation in an ensemble of qubits kicks in. Note that the existence of entangled dark states has
been recently predicted in \cite{gray2015,gray2016} by calculating the
concurrence with the master equation in the Lindblad approximation. 

In Secs.~III and IV the quantum evolution of two qubits with arbitrary vacuum Rabi frequencies and frequency detunings from the cavity mode are described. The formation of dark states decoupled from the cavity mode due to destructive quantum interference is predicted. In Sec.~V we calculate the emission spectra for an ensemble of $N$ qubits in a leaky cavity. 

Finally, in Sec.~VI, we solve for the evolution of the  electron systems in which  transition frequencies may differ by a large
value, much larger than the homogeneous linewidth of an individual emitter
and the linewidth of a cavity mode. This introduces an additional
effective decoherence for an ensemble of qubits, in addition to the coupling
of electrons and photons to their dissipative reservoirs. At the same time,
different transition frequencies introduce extra degrees of freedom and add
more functionality to the system. Examples of such systems include
semiconductor quantum wells, quantum dots, or defects in crystals.

Since all results in the paper are in the analytic form and the plots are normalized by the single-qubit vacuum Rabi frequency, here we list typical values of the parameters in experimental solid-state nanophotonic systems that determine the strength of light-matter coupling and relaxation rates. To determine if the strong coupling regime and quantum entanglement in the electron-photon system can be achieved,  one has to compare the relaxation rates with the characteristic Rabi frequency which determines the oscillations of energy between the cavity field and the qubits. The strong-coupling conditions for an ensemble of $N$ qubits are easier to implement since the effective Rabi frequency scales as $\sqrt{N}$. At the same time, placing a large number of qubits in a nanocavity may present a challenge by itself as individual quantum emitters  will experience nonuniform field distribution of a cavity mode and the coupling between them is difficult to control and may lead to additional decoherence. 

The parameters mentioned below were obtained from recent experimental reports and reviews such as \cite{thorma2015, lodahl2015, degen2017,  dovzhenko2018, bitton2019, todorov2010, forndiaz2017, kono2019, chikkaraddy2016,benz2016, park2016, pelton2018, gross2018, park2019, lukin2016,deppe,reithmaier,sercel2019, fieramosca2019} . This paper does not attempt to provide a comprehensive overview of the rapidly growing literature on the subject.

In electron-based quantum emitters the largest oscillator strengths in the visible/near-infrared range have been observed for excitons in organic molecules, followed by perovskites and more conventional inorganic semiconductor quantum dots. The typical variation of the dipole matrix element of the optical transition which enters the Rabi frequency is from tens of nm to a few Angstrom (in units of the electron charge). Within the same kind of a system, the dipole moment grows with increasing wavelength. The relaxation times are strongly temperature and material quality-dependent, varying from tens or hundreds of ps for electric-dipole allowed interband transitions in quantum dots at 4 K to the tens and hundreds of $\mu$s for spin qubits based on defects in semiconductors and diamond at mK temperatures. Dipole-forbidden optical transitions, e.g. dark excitons in quantum dots, can have the lifetime up to milliseconds, but their coupling to light could be too weak to reach the strong coupling regime.  At room temperature the decoherence rates for the optical transitions are in the tens of meV range.

The photon decay times are longest for dielectric micro- and nano-cavities: photonic crystal cavities, nanopillars, distributed Bragg reflector mirrors, microdisk whispering gallery mode cavities, etc. Their quality factors are typically between $10^3 - 10^7$, corresponding to photon lifetimes from sub-ns to $\mu$s range. However, the field localization in the dielectric cavities is diffraction-limited, which limits the attainable single-qubit vacuum Rabi frequency values to hundreds of $\mu$eV \cite{deppe,reithmaier}. The effective decay rate of the eigenstates (e.g. exciton-polaritons)  in dielectric cavity QED systems is typically limited by the relaxation in the fermion quantum emitter subsystem. 

In plasmonic cavities, field localization on a nm and even sub-nm scale has been achieved, but the photon decay time is in the ps or even fs range and therefore, the photon losses dominate the overall decoherence rate. Still, when it comes to strong coupling at room temperature to a single quantum emitter such as a single molecule or a quantum dot, the approach utilizing plasmonic nanocavities has seen more success so far. In these systems the Rabi splitting of the order of 100-200 meV has been observed \cite{chikkaraddy2016,benz2016, park2016, pelton2018, gross2018, park2019}.

\section{ N identical qubits}

Consider first a textbook problem of $N$ identical two-level systems with
states $\left\vert 0_{j}\right\rangle $ and $\left\vert 1_{j}\right\rangle $%
, where $j=1,...N$, with energy levels $0$ and $W$. We introduce fermionic
operators of annihilation and creation of an excited state $\left\vert
1_{j}\right\rangle $,
\begin{equation}
\hat{\sigma}_{j}=\left\vert 0_{j}\right\rangle \left\langle 1_{j}\right\vert
,\ \ \ \hat{\sigma}_{j}^{\dagger }=\left\vert 1_{j}\right\rangle
\left\langle 0_{j}\right\vert ,  \label{sigma j}
\end{equation}%
the dipole moment operator, 
\begin{equation}
\mathbf{\hat{d}}=\sum_{j=1}^{N} \left( \mathbf{d} \hat{\sigma}_{j}^{\dagger }+ \mathbf{d^*} 
\hat{\sigma}_{j}\right),  \label{dmo}
\end{equation}
and the Hamiltonian for all qubits,
\begin{equation}
\hat{H}_{a}=W\sum_{j=1}^{N}\hat{\sigma}_{j}^{\dagger }\hat{\sigma}_{j}.
\label{Ha}
\end{equation}%
Here $\mathbf{d=}\left\langle 1_{j}\right\vert \mathbf{\hat{d}}\left\vert
0_{j}\right\rangle $ is the same for all $j$. Our $N$-qubit system interacts
with a single-mode field 
\begin{equation}
\mathbf{\hat{E}}=\mathbf{E}\left( \mathbf{r}\right) \hat{c}+\mathbf{E}^{\ast
}\left( \mathbf{r}\right) \hat{c}^{\dagger },  \label{E}
\end{equation}%
where $\hat{c}$ and $\hat{c}^{\dagger }$ are standard annihilation and
creation operators for bosonic Fock states. 

The function $\mathbf{E}\left( \mathbf{r}\right) $ determines the spatial
structure of the electric field in a cavity. It is normalized as \cite{Tokman2016}
\begin{equation}
\int_{V}\frac{\partial \left[ \omega ^{2}\varepsilon \left( \omega ,\mathbf{r%
}\right) \right] }{\omega \partial \omega }\mathbf{E}^{\ast }\left( \mathbf{r%
}\right) \mathbf{E}\left( \mathbf{r}\right) d^{3}r=4\pi \hbar \omega
\label{nc-const}
\end{equation}%
to preserve the standard form of the field Hamiltonian,
\begin{equation}
\hat{H}_{em}=\hbar \omega \left( \hat{c}^{\dagger }\hat{c}+\frac{1}{2}\right). 
\label{HEM}
\end{equation}%

The relation between the modal frequency $\omega $ and the function $\mathbf{%
E}\left( \mathbf{r}\right) $ can be found by solving the boundary-value
problem of the classical electrodynamics. Here $V$ is a quantization volume
and $\varepsilon \left( \omega ,\mathbf{r}\right) $ is the dielectric
function of a dispersive medium that fills the cavity.

The total Hamiltonian after adding interaction with the field is
\begin{equation}
\hat{H}_{a+em}=\hat{H}_{a}+\hat{H}_{em}-\mathbf{\hat{d}}\cdot \left( \mathbf{%
E}\hat{c}+\mathbf{E}^{\ast }\hat{c}^{\dagger }\right) ,  \label{Hema}
\end{equation}
\begin{equation}
\hat{H}=\hbar \omega \left( \hat{c}^{\dagger }\hat{c}+\frac{1}{2}\right)
+\sum_{j=1}^{N}\left[ W\hat{\sigma}_{j}^{\dagger }\hat{\sigma}_{j}-\left(
\mathbf{d}\cdot \mathbf{E}\hat{\sigma}_{j}^{\dagger }\hat{c}+\mathbf{d}%
^{\ast }\cdot \mathbf{E}\hat{\sigma}_{j}\hat{c}+\mathbf{d}\cdot \mathbf{E}%
^{\ast }\hat{\sigma}_{j}^{\dagger }\hat{c}^{\dagger }+\mathbf{d}^{\ast
}\cdot \mathbf{E}^{\ast }\hat{\sigma}_{j}\hat{c}^{\dagger }\right) \right] ,
\label{H}
\end{equation}
where we assumed that the field amplitude is the same at the position of all
qubits. If the field frequency is close to resonance,
\begin{equation}
\frac{W}{\hbar }\approx \omega  \label{rc}
\end{equation}%
we can use the RWA Hamiltonian in which counter-rotating terms are dropped:%
\begin{equation}
\hat{H}=\hbar \omega \left( \hat{c}^{\dagger }\hat{c}+\frac{1}{2}\right)
+\sum_{j=1}^{N}\left[ W\hat{\sigma}_{j}^{\dagger }\hat{\sigma}_{j}-\left(
\mathbf{d}\cdot \mathbf{E}\hat{\sigma}_{j}^{\dagger }\hat{c}+\mathbf{d}%
^{\ast }\cdot \mathbf{E}^{\ast }\hat{\sigma}_{j}\hat{c}^{\dagger }\right) %
\right] ,  \label{RWAH}
\end{equation}

The Schr\"{o}dinger equation with the Hamiltonian (\ref{RWAH}) is separable into low-dimensional blocks and solvable analytically for any excitation energy and for an arbitrary multiphoton state; see the Appendix A for a general treatment. However, the generation of nonclassical multiphoton states is still an experimental challenge and there is no compelling reason in presenting more general but also more cumbersome formulas for multiphoton excitations. That is why we choose the initial conditions for which the transitions to states beyond single-photon excitations are forbidden by energy conservation. These are most popular states widely used in quantum information applications, including for example Bell states and their generalizations to many-qubit systems. Then the states that can be reached as a result of evolution of the system 
 include the ground state $\left\vert 0\right\rangle
\Pi_{j=1}^{N}\left\vert 0_{j}\right\rangle $ and the states with energies
close to the single-photon energy:
\begin{equation}
\Psi =C_{00}\left\vert 0\right\rangle \Pi _{j=1}^{N}\left\vert
0_{j}\right\rangle +C_{10}\left\vert 1\right\rangle \Pi _{j=1}^{N}\left\vert
0_{j}\right\rangle +\sum_{j=1}^{N}C_{0j}\left\vert 0\right\rangle \left\vert
1_{j}\right\rangle \Pi _{m\neq j}^{N}\left\vert 0_{m}\right\rangle.
\label{Psi}
\end{equation}

Substituting the state vector (\ref{Psi}) into Eq.~(\ref{RWAH}) leads to the
equations for coefficients,
\begin{equation}
\dot{C}_{00}+i\frac{\omega }{2}C_{00}=0,  \label{C00 dot}
\end{equation}%
\begin{equation}
\dot{C}_{10}+i \frac{3 \omega}{2} C_{10}-i\Omega _{R}^{\ast
}\sum_{j=1}^{N}C_{0j}=0,  \label{C10 dot}
\end{equation}%
\begin{equation}
\dot{C}_{0j}+i\left( \frac{\omega}{2}+\frac{W}{\hbar }\right) C_{0j}-i\Omega
_{R}C_{10}=0,  \label{C0j dot}
\end{equation}%
where the Rabi frequency $\Omega _{R}=\frac{\mathbf{d\cdot E}}{\hbar }$. Let $%
\sum_{j=1}^{N}C_{0j}=F$. Then the equation for $C_{00}(t)$ is decoupled, 
\begin{equation}
C_{00}(t) = C_{00}(0) e^{-i \frac{\omega}{2} t};
\end{equation}
and the other two equations,
\begin{equation}
\dot{C}_{10}+i \frac{3 \omega }{2} C_{10}-i\Omega _{R}^{\ast }F=0,
\label{C10 dot F}
\end{equation}
\begin{equation}
\dot{F}+i\left( \frac{\omega}{2}+\frac{W}{\hbar }\right) F-iN\Omega
_{R}C_{10}=0,  \label{F dot}
\end{equation}
can be easily solved. Indeed, after making the substitution
\begin{equation*}
\left(
\begin{array}{c}
C_{10} \\
F%
\end{array}%
\right) =\left(
\begin{array}{c}
G_{10} \\
G_{F}%
\end{array}%
\right) e^{-i \frac{3 \omega}{2} t},
\end{equation*}
we obtain
\begin{equation}
\dot{G}_{10}-i\Omega _{R}^{\ast }G_{F}=0,  \label{G10 dot}
\end{equation}%
\begin{equation}
\dot{G}_{F}+i\Delta G_{F}-i\Omega _{R}NG_{10}=0,  \label{GF dot}
\end{equation}%
where $\Delta =\frac{W}{\hbar }-\omega .$

This leads to the following equation for the variable $G_{10}$ which
describes the quantum state of the field,
\begin{equation}
\ddot{G}_{10} + i \Delta \dot{G}_{10} +N |\Omega _{R}|^2 G_{10} = 0,
\label{G10 ddot}
\end{equation}
which looks exactly like the one for a single qubit, but with an $N$-qubit
Rabi frequency obtained by $\left\vert \Omega _{R}\right\vert
^{2}\Longrightarrow N\left\vert \Omega _{R}\right\vert ^{2}$. It has the
solution
\begin{equation*}
G_{10} = A_1 e^{\Gamma_1 t} + A_2 e^{\Gamma_2 t},
\end{equation*}
where the constants $A_{1,2}$ are determined by initial conditions and
\begin{equation}
\Gamma _{1,2}=-i\frac{\Delta }{2}\pm i\sqrt{\frac{\Delta ^{2}}{4}%
+N\left\vert \Omega _{R}\right\vert ^{2}}.  \label{Gamma}
\end{equation}

Then the solution for coefficients $C_{0j}$ can be obtained by substituting
the expression for $C_{10}$ into Eq.~(\ref{C0j dot}),
\begin{equation}
C_{0j}(t) = e^{-i \left( \frac{3 \omega}{2} + \Delta\right) t} \left(
C_{0j}(0) + i \Omega_R \int_0^t e^{i \Delta \tau} G_{10}(\tau) \,d\tau
\right).
\end{equation}

Therefore, a system of $N$ identical qubits responds to the EM field as 
one qubit with a ``giant'' dipole moment $\propto \sqrt{N}$. At the same time, the qubits
themselves demonstrate a much more complex dynamics. Their interaction with
a quantum field leads to their entanglement not only with the field but also
with each other; see, e.g., \cite{TMD2015}. These quantum correlations
within the ensemble of qubits can profoundly alter the evolution of the
whole system. We illustrate it with a few examples.

First, consider our system in the factorized initial state $\Psi(0) =
\left\vert 1\right\rangle \Pi_{j=1}^{N}\left\vert 0_{j}\right\rangle $, in
which the photon mode is excited, all qubits are in their ground states, and
there is no entanglement between any degrees of freedom. Considering for
simplicity the case $\Delta = 0$, the subsequent evolution of the state is
given by
\begin{equation}
\Psi = e^{-i \frac{3 \omega}{2} t} \left[ \cos\left(\sqrt{N} |\Omega _{R}| t
\right) \left\vert 1\right\rangle \Pi _{j=1}^{N}\left\vert
0_{j}\right\rangle + \frac{i e^{i \theta}}{\sqrt{N}} \sin\left(\sqrt{N} |
\Omega _{R}| t \right) \sum_{j = 1}^N \left\vert 0\right\rangle \left\vert
1_j\right\rangle \Pi _{m\neq j}^{N}\left\vert 0_{m}\right\rangle \right],
\label{Psi2}
\end{equation}
where $\theta = \mathrm{Arg}[\Omega_R]$. At the moments of time $\sqrt{N}
|\Omega_R| t = \frac{\pi}{2} + \pi M$ the qubits are entangled with each
other in a Bell-type state but not entangled with the field, whereas at an
arbitrary moment of time, there are entangled both with each other and with
the field. The excitation energy periodically oscillates between the cavity
mode and an entangled state of the qubits. If the system is open, the
excitation will dissipate through relaxation in both photon and qubit
subsystems.

Now consider a different factorized initial state $\Psi(0) = \left\vert
0\right\rangle \left\vert 1_1\right\rangle \Pi_{j=2}^{N}\left\vert
0_{j}\right\rangle $, in which only one qubit is excited, for definiteness
the one with $j = 1$. This implies that qubits can be individually
addressed, for example by a beam of light or a near-field nanotip with
excitation radius much smaller than the distance between the qubits.

Taking again $\Delta = 0$ for simplicity, we obtain the solution
\begin{equation}
\begin{array}{c}
\Psi = e^{-i \frac{3 \omega}{2} t} \left[ \frac{i e^{-i \theta}}{\sqrt{N}}
\sin\left(\sqrt{N} | \Omega _{R}| t \right) \left\vert 1\right\rangle \Pi
_{j=1}^{N}\left\vert 0_{j}\right\rangle + \left( 1- \frac{1 - \cos\left(%
\sqrt{N} |\Omega _{R}| t \right)}{N} \right) \left\vert 0\right\rangle
\left\vert 1_1 \right\rangle \Pi _{j=2}^{N} \left\vert 0_{j}\right\rangle
\right. \\
\left. - \frac{1 - \cos\left(\sqrt{N} |\Omega _{R}| t \right)}{N} \sum_{j =
2}^N \left\vert 0\right\rangle \left\vert 1_j \right\rangle \Pi _{m\neq
j}^{N}\left\vert 0_{m}\right\rangle \right],%
\end{array}
\label{Psi3}
\end{equation}
It follows from the time-dependent coefficients in Eq.~(\ref{Psi3}) that for
$N \gg 1$ the occupation of the initial state given by the second term on
the right-hand side of Eq.~(\ref{Psi3}) oscillates around 1 with a small
amplitude $\frac{1}{N}$, whereas the occupations of all other states
oscillate around zero with equally small amplitudes. This means that the
probability of the energy transfer from the excited qubit to the field or to
an ensemble of ground-state qubits is about $N$ times smaller than the
probability that the initial state is preserved, which scales as $\left\vert
1- \frac{1 - \cos\left(\sqrt{N} |\Omega _{R}| t \right)}{N} \right\vert^2
\geq \left( 1-\frac{2}{N} \right)^2$.

This result holds true whether or not we know which particular qubit is
excited, as long as the excitation radius remains much smaller than the
distance between the qubits. Contrary to famous double-slit experiments, for
the qubit excitation the ``which qubit'' information constitutes a classical
uncertainty. Indeed, imagine that we illuminate the qubits with a tightly
confined beam from the far field or with a near field of the tip. The answer
to the question "which qubit got excited" depends on the spatial mode
structure which is determined by classical Maxwell's equations. Therefore,
the created state will have the structure $\Psi(0) = \left\vert
0\right\rangle \left\vert 1_j\right\rangle \Pi_{q \neq j}^{N}\left\vert
0_{q}\right\rangle $, where the classical uncertainty in the value of the
qubit number $j$ does not give rise to the entangled quantum state of the
type $\Psi(0) = \frac{1}{\sqrt{N}} \sum_{j=1}^N \left\vert 0\right\rangle
\left\vert 1_j\right\rangle \Pi_{q \neq j}^{N}\left\vert 0_{q}\right\rangle $
(as quantum uncertainty would).

This reasoning can be generalized to the situation in which a group of $J$
qubits fits inside the excitation radius, where $J \ll N$. We show in
Sec.~IV below that if a group of $J \ll N$ qubits is initially excited,
their deexcitation will be suppressed too and the initial state will be
preserved with probability $\sim 1-\frac{J}{N}$. Here again, the uncertainty
of the type ``which group is illuminated'' is classical. At the same time,
for a beam carrying single-photon energy there is a quantum uncertainty
``which qubit inside the group will get excited''.

%Figure ?? shows the probabilities of states with excited photon $|C_{10}|^2$
% and excited qubit $|C_{0j}|^2$ as a function of time. Small amplitudes of
% the oscillations illustrate the effect of the suppression of the excitation
% transfer.

The presence of $N>1$ qubits allows the existence of entangled qubit states
that are completely decoupled from the cavity field. Indeed, for the initial
condition $C_{10}=0$ we can find any number of entangled qubit states
satisfying the equation $\left\vert C_{00}\right\vert
^{2}+\sum_{j=1}^{N}\left\vert C_{0j}\right\vert ^{2}=1$, $%
\sum_{j=1}^{N}C_{0j}=0$, i.e.~the states that are decoupled from the field.
This solution is of the type
\begin{equation*}
\left(
\begin{array}{c}
C_{10} \\
\vdots \\
C_{0j} \\
\vdots%
\end{array}%
\right) \propto e^{-i \frac{3 \omega}{2} t} \left(
\begin{array}{c}
0 \\
\vdots \\
A_{0j}e^{-i\Delta t} \\
\vdots%
\end{array}%
\right) ,\ \ \ \sum_{j=1}^{N}A_{0j}=0.
\end{equation*}

The partial and complete decoupling of qubits from the field can be
illustrated by considering initial Bell-type states of the two qubits:
\begin{equation}
\Psi_{\pm}(0) = \left\vert 0\right\rangle \frac{\left\vert 1_1\right\rangle
\left\vert 0_2 \right\rangle \pm \left\vert 0_1\right\rangle \left\vert 1_2
\right\rangle}{\sqrt{2}} \Pi _{j=3}^{N}\left\vert 0_{j}\right\rangle.
\label{bell}
\end{equation}
Consider again the exact resonance for simplicity, $\Delta = 0$. If the
field is not excited initially, i.e., $C_{10}(0) = 0$, the solution for the
complex probability amplitudes takes the form
\begin{equation}
C_{10}(t) = i e^{-i \theta -i \frac{3 \omega}{2} t} \frac{F(0)}{\sqrt{N}}
\sin\left(\sqrt{N} |\Omega _{R}| t \right),  \label{bell2}
\end{equation}
\begin{equation}
C_{0j}(t) = e^{-i \frac{3 \omega}{2} t} \left( C_{0j}(0) - \frac{F(0)}{N} %
\left[ 1 - \cos\left(\sqrt{N} |\Omega _{R}| t \right) \right] \right),
\label{bell3}
\end{equation}
where $F(0) = \sum_{j = 1}^N C_{0j}(0)$. In our case we have $C_{01}(0) =
\frac{1}{\sqrt{2}}$ and $C_{02}(0) = \pm \frac{1}{\sqrt{2}}$ for $%
\Psi_{\pm}(0)$ respectively; $C_{0;j>2} = 0$. Therefore, for $\Psi_{(-)}(0)$
we obtain $F(0) = 0$, i.e., $\Psi_{(-)}(t) = e^{-i \frac{3 \omega}{2} t}
\Psi_{(-)}(0)$ and the state of two qubits is completely uncoupled from the
field.

For $\Psi_{(+)}(0)$ we obtain $F(0) = \sqrt{2}$ and
\begin{equation}
C_{10}(t) = i e^{-i \theta -i \frac{3 \omega}{2} t} \frac{\sqrt{2})}{\sqrt{N}%
} \sin\left(\sqrt{N} |\Omega _{R}| t \right),  \label{bell4}
\end{equation}
\begin{equation}
C_{0;1,2}(t) = e^{-i \frac{3 \omega}{2} t} C_{0;1,2}(0) \left( 1 - \frac{2}{N%
} \left[ 1 - \cos\left(\sqrt{N} |\Omega _{R}| t \right) \right] \right),
\label{bell5}
\end{equation}
\begin{equation}
C_{0;j>2}(t) = C_{0;1,2}(0) \frac{2}{N} \left[ 1 - \cos\left(\sqrt{N}
|\Omega _{R}| t \right) \right].  \label{bell6}
\end{equation}
Obviously, for $\Psi_{(+)}(0)$ the probability amplitudes are changing but
the change is small as long as $\frac{2}{N} \ll 1$.

To summarize, if no qubits are excited in the initial state, their ensemble
interacts with a photon as one giant qubit. If the system is in the 
initial state in which one qubit or a small number of qubits are
initially excited, the energy transfer of its excitation to other qubits and
the photon mode will be slowed down by a factor of $N$. This will
significantly slow down the relaxation of the initial excitation if the
relaxation in the qubit subsystem is slower than that for the cavity mode.
Moreover, there are entangled qubit states that are completely decoupled
from the cavity field. For a solid state system with large $N$, the
excitation and control of such or any other specified entangled multi-qubit
states can be tricky. However, as we show below, the EM field dissipation may drive the multi-qubit system towards such states.

The monolithic solid-state systems where a very large $N$ can be reached
include, for example, intersubband transitions in doped quantum wells (QWs),
excitons in semiconductor QWs or 2D semiconductors, and a 2D electron gas in
a strong magnetic field. In these systems, strong and even ultra-strong
coupling to a cavity field has been demonstrated, although so far with a
classical EM field; e.g., \cite{todorov2010,kono2019}. Once the quantized
field regime is reached, one can imagine coupling several N-qubit systems to
a common cavity field, e.g. a multi-quantum well system or excitons in
different valleys of 2D semiconductors \cite{TMD2015}. The control of
specific single-qubit excitations in the monolithic systems is a challenge.
If such a control is desirable, it could be easier to deal with a system of
spatially isolated qubits in a cavity, e.g. quantum dots or molecular
qubits, as proposed in \cite{gray2015}.

%%%%%%%%%%%%%%%%%%%%%%%%%%%%%%%%%%%%%%%%%%%%%%%%%%%%%%%%%%%%%%%%%%%%

\section{Two qubits with different transition frequencies}

When the qubits are not identical, this opens more possibilities for quantum
state manipulation but also makes the dynamics more complex. The
conceptually simplest system includes two qubits which have different
transition frequencies. Here we describe their dissipationless dynamics
whereas the effects of dissipation are included in Sec.~IV.

Assume that both transition frequencies are close enough to the field
frequency, so that we can still use the RWA:
\begin{equation}
\hat{H}=\hbar \omega \left( \hat{c}^{\dagger }\hat{c}+\frac{1}{2}\right)
+\sum_{j=1,2}\left[ W_{j}\hat{\sigma}_{j}^{\dagger }\hat{\sigma}_{j}-\hbar
\Omega _{Rj}\left( \hat{\sigma}_{j}^{\dagger }\hat{c}+\hat{\sigma}_{j}\hat{c}%
^{\dagger }\right) \right] ,  \label{Hamiltonian 2 qubits}
\end{equation}%
where $\Omega _{R1,2}=\frac{\mathbf{d}_{1,2}\mathbf{\cdot E}}{\hbar }$.
However, the frequency detuning values can be arbitrary with respect to Rabi
frequencies. Consider only the ground state and lowest excited state with
excitation energy close to single photon energy (i.e. $n=1$). The wave
function including only those states is
\begin{equation}
\Psi =C_{000}\left\vert 0\right\rangle \left\vert 0\right\rangle \left\vert
0\right\rangle +C_{100}\left\vert 1\right\rangle \left\vert 0\right\rangle
\left\vert 0\right\rangle +C_{010}\left\vert 0\right\rangle \left\vert
1\right\rangle \left\vert 0\right\rangle +C_{001}\left\vert 0\right\rangle
\left\vert 0\right\rangle \left\vert 1\right\rangle.  \label{psi}
\end{equation}
Here the order of the terms is always $\left\vert photon\right\rangle
\left\vert atomN1\right\rangle \left\vert atomN2\right\rangle $.

From the Schr\"{o}dinger equation we obtain
\begin{equation}
\dot{C}_{000}+i\frac{\omega }{2}C_{000}=0;  \label{c000 dot}
\end{equation}%
\begin{equation}
\dot{C}_{100}+i\omega \frac{3}{2}C_{100}-i\left( \Omega
_{R1}^*C_{010}+\Omega _{R2}^*C_{001}\right) =0,  \label{c100 dot}
\end{equation}%
\begin{equation}
\dot{C}_{010}+i\left( \omega \frac{1}{2}+\frac{W_{1}}{\hbar }\right)
C_{010}-i\Omega _{R1}C_{100}=0,  \label{c010 dot}
\end{equation}%
\begin{equation}
\dot{C}_{001}+i\left( \omega \frac{1}{2}+\frac{W_{2}}{\hbar }\right)
C_{001}-i\Omega _{R2}C_{100}=0,  \label{c001 dot}
\end{equation}%
One can eliminate high frequencies with $\left(
\begin{array}{c}
C_{100} \\
C_{010} \\
C_{001}%
\end{array}%
\right) =\left(
\begin{array}{c}
G_{100} \\
G_{010} \\
G_{001}%
\end{array}%
\right) e^{-i\omega \frac{3}{2}t},$ which gives
\begin{equation}
\dot{G}_{100}-i\left( \Omega _{R1}^*G_{010}+\Omega _{R2}^*G_{001}\right) =0,
\label{G100 eq}
\end{equation}%
\begin{equation}
\dot{G}_{010}+i\Delta _{1}G_{010}-i\Omega _{R1}G_{100}=0,  \label{G010 eq}
\end{equation}%
\begin{equation}
\dot{G}_{001}+i\Delta _{2}G_{001}-i\Omega _{R2}G_{100}=0,  \label{G001 eq}
\end{equation}%
where $\Delta _{1,2}=\frac{W_{1,2}}{\hbar }-\omega .$ The solution for
coefficients can be found as $\left(
\begin{array}{c}
G_{100} \\
G_{010} \\
G_{001}%
\end{array}%
\right) \propto e^{\Gamma t};$

\begin{equation*}
\left(
\begin{array}{ccc}
\Gamma & -i\Omega _{R1}^* & -i\Omega _{R2}^* \\
-i\Omega _{R1} & \Gamma +i\Delta _{1} & 0 \\
-i\Omega _{R2} & 0 & \Gamma +i\Delta _{2}%
\end{array}%
\right) \left(
\begin{array}{c}
G_{100} \\
G_{010} \\
G_{001}%
\end{array}%
\right) =0,
\end{equation*}
which gives an equation for eigenvalues,
\begin{equation}
\Gamma \left( \Gamma +i\Delta _{1}\right) \left( \Gamma +i\Delta _{2}\right)
+|\Omega _{R1}|^{2}\left( \Gamma +i\Delta _{2}\right) +|\Omega
_{R2}|^{2}\left( \Gamma +i\Delta _{1}\right) =0  \label{eq for gamma}
\end{equation}%
This is a cubic equation which can be solved analytically or on a computer.

\begin{figure}[htb]
\includegraphics[width=0.5\linewidth]{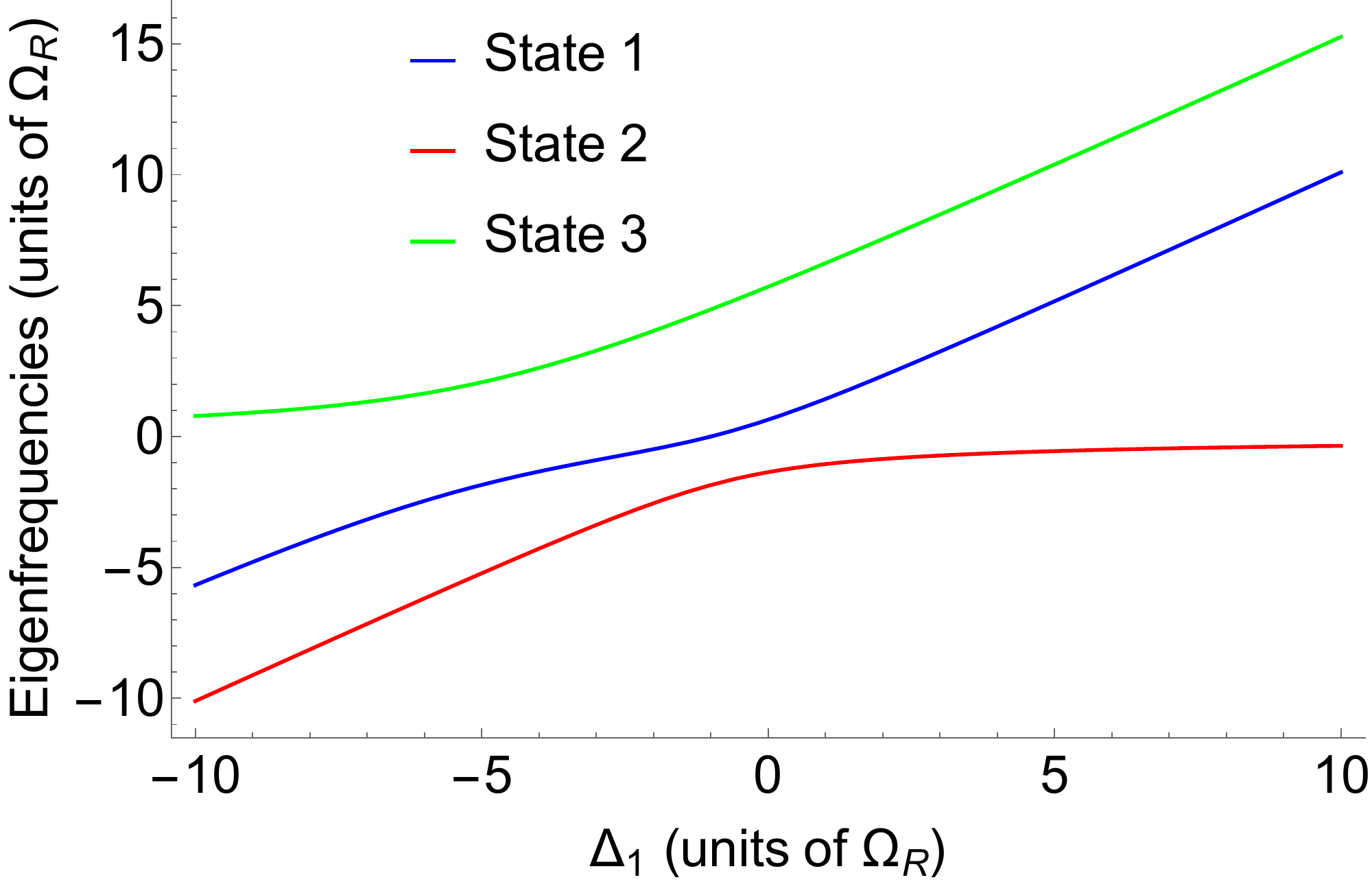}
\caption{ Frequency eigenvalues from Eq.~(\protect\ref{eq for gamma}) as a
function of detuning $\Delta_1$, when $|\Omega_{R1}| = \Omega_{R}$, $%
|\Omega_{R2}| = 2 \Omega_{R}$, and $\Delta _{2}=\Delta _{1} + 5 \Omega_{R}$.
All frequencies are in units of $\Omega _{R}$.}
\label{fig1}
\end{figure}

\begin{figure}[htb]
\includegraphics[width=0.5\linewidth]{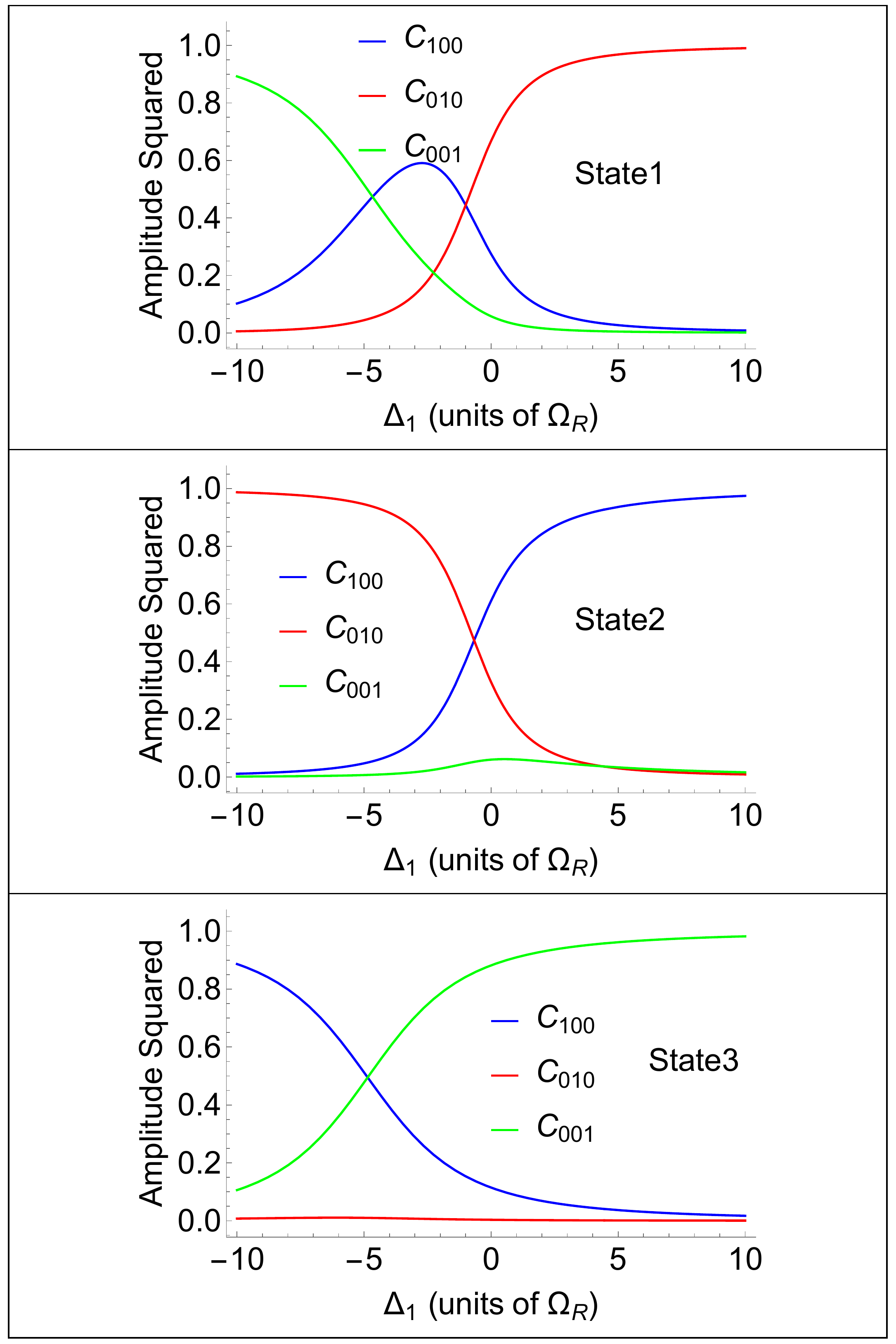}
\caption{Amplitudes squared of the eigenstates from Eq.~(\protect\ref{eq for
gamma}) as a function of detuning $\Delta_1$, when $|\Omega_{R1}| =
\Omega_{R}$, $|\Omega_{R2}| = 2 \Omega_{R}$, and $\Delta _{2}=\Delta _{1} +
5 \Omega_{R}$. }
\label{fig2}
\end{figure}

Figures 1 and 2 show one example of the solution for eigenenergies and
eigenstates, respectively, when $|\Omega_{R1}| = \Omega_{R}$, $|\Omega_{R2}|
= 2 \Omega_{R}$, and $\Delta _{2}=\Delta _{1} + 5 \Omega_{R}$. Two
anticrossing points are visible in the spectra in Fig.~1. As is clear from
Fig.~2, by controlling the initial conditions for a given eigenstate and
tuning through resonances with a cavity mode one can arrive at an arbitrary entangled state or
transfer the excitation from one qubit to another or from any of the qubits
to the photon state.

If $\Delta _{1}=\Delta _{2}=\Delta $, we obtain
\begin{equation}
\left( \Gamma +i\Delta \right) \left( \Gamma ^{2}+i\Delta \Gamma +|\Omega
_{R1}|^{2}+|\Omega _{R2}|^{2}\right) =0.  \label{gamma equ}
\end{equation}

In addition to the solutions of $\ \Gamma ^{2}+i\Delta \Gamma +|\Omega
_{R1}|^{2}+|\Omega _{R2}|^{2}=0$ denoted as states $a$ and $b$ in Figs.~3
and 4 below, we have a root $\Gamma =-i\Delta $ for any $\Omega _{R1,2}$.
This solution is a dark state which does not couple to the field:
\begin{equation}
\left(
\begin{array}{c}
C_{100} \\
C_{010} \\
C_{001}%
\end{array}%
\right) \propto \left(
\begin{array}{c}
0 \\
1 \\
-\frac{\Omega _{R1}^*}{\Omega _{R2}^*}%
\end{array}%
\right) .  \label{dark}
\end{equation}

\begin{figure}[htb]
\includegraphics[width=0.5\linewidth]{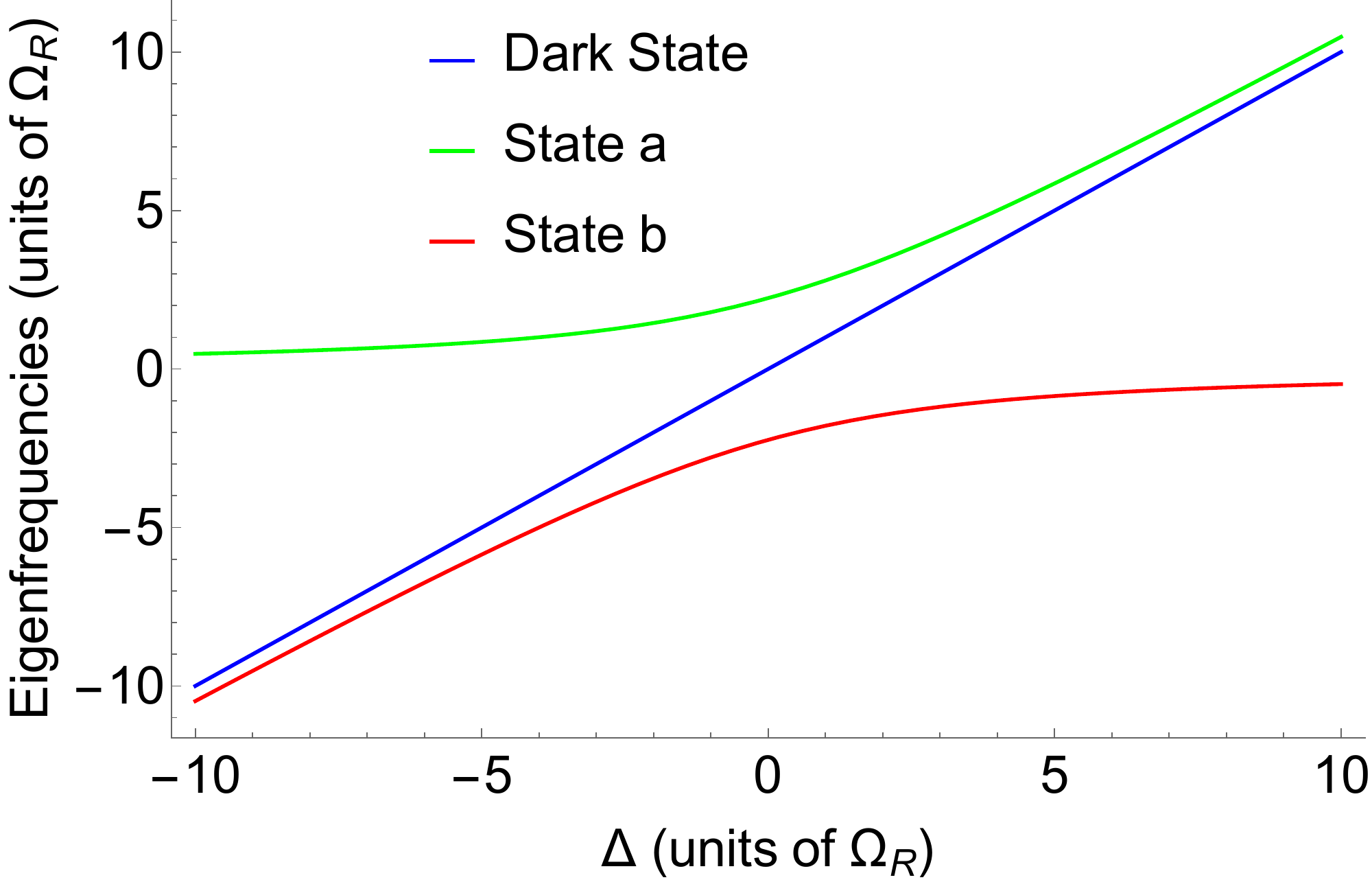}
\caption{ Frequency eigenvalues from Eq.~(\protect\ref{eq for gamma}) for $%
\Delta _{1}=\Delta _{2}=\Delta $ as a function of detuning $\Delta$, when $%
|\Omega_{R1}| = \Omega_{R}$ and $|\Omega_{R2}| = 2 \Omega_{R}$. The dark
state which is decoupled from the cavity mode at any detuning is defined in
Eq.~\protect\ref{dark}). }
\label{fig3}
\end{figure}

\begin{figure}[htb]
\includegraphics[width=0.5\linewidth]{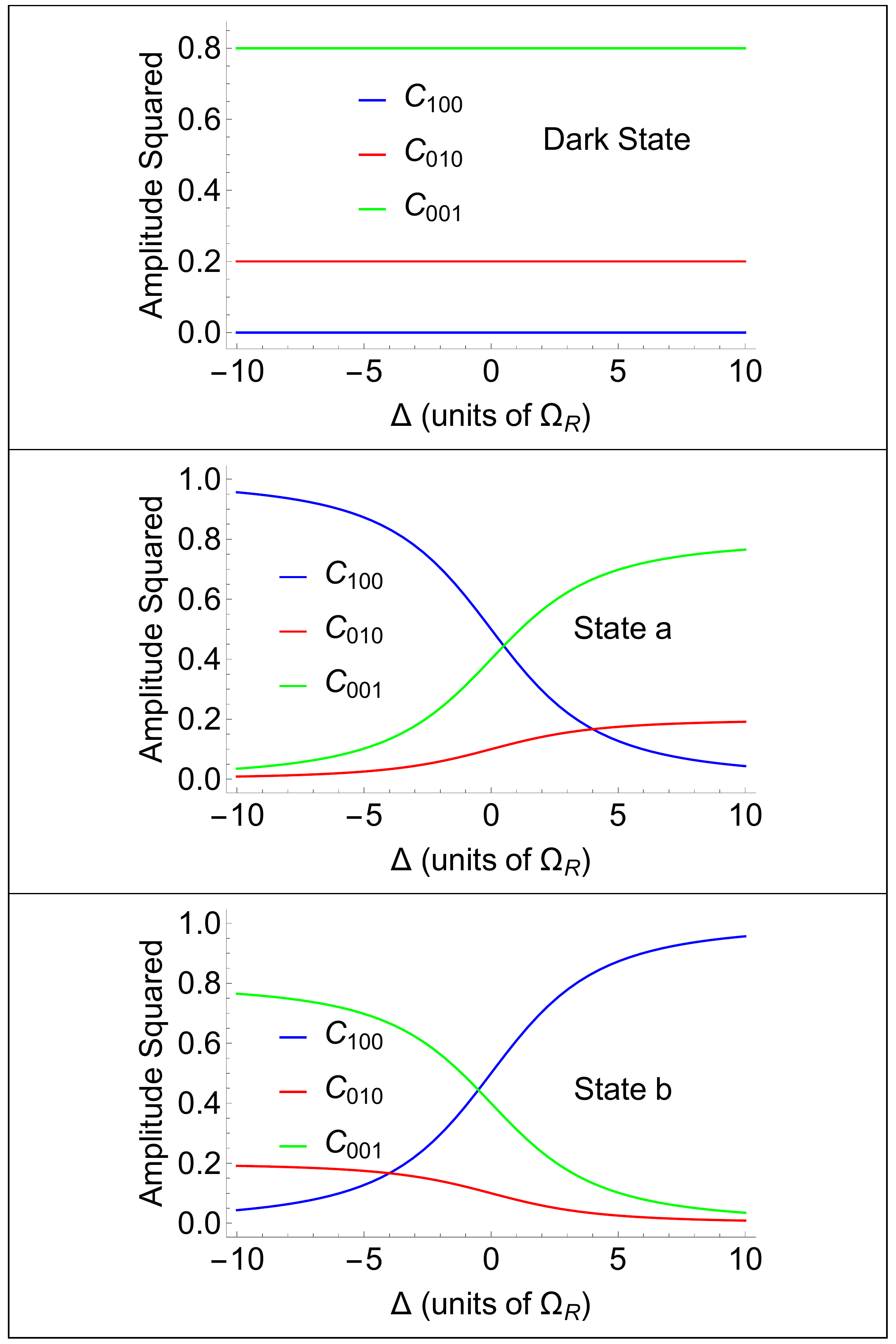}
\caption{Amplitudes squared of the eigenstates from Eq.~(\protect\ref{eq for
gamma}) for $\Delta _{1}=\Delta _{2}=\Delta $ as a function of detuning $%
\Delta$, when $\Omega_{R1} = \Omega_{R}$ and $\Omega_{R2} = 2 \Omega_{R}$.
The dark state which is decoupled from the cavity mode at any detuning is
defined in Eq.~\protect\ref{dark}). }
\label{fig4}
\end{figure}

Figures 3 and 4 illustrate this example, showing eigenenergies and
eigenstates for $\Delta _{1}=\Delta _{2}=\Delta $. Clearly, one state
denoted as Dark state on the figures is decoupled from light at any
detunings whereas the two bright states $a$ and $b$ interact strongly
through coupling to a cavity mode. By changing the initial conditions one
can store the excitation in a dark state and read it out on demand, of
course within the limits imposed by decoherence.

As a final example, consider the case $\Delta _{1}=0$, $\Delta _{2}=\Delta $%
, which gives
\begin{equation}
|\Omega _{R2}|^{2}\Gamma +\left( |\Omega _{R1}|^{2}+\Gamma ^{2}\right)
\left( \Gamma +i\Delta \right) =0.  \label{gamma equations}
\end{equation}
In particular, when $\left\vert \Delta \right\vert \gg |\Omega _{R1,2}|$
there is an approximate analytic solution $\Gamma _{1}\approx -i\Delta $, $%
\Gamma _{2,3}\approx \pm |\Omega _{R1}|$ which means that the field couples
only to a resonant qubit.

%%%%%%%%%%%%%%%%%%%%%%%%%%%%%%%%%

\section{Dynamics of multi-qubit systems in the presence of dissipation}

\subsection{The stochastic equation of evolution for the state vector}

A standard way to include the effects of dissipation is based on the master
equation for the density matrix $\hat{\rho}$ of the system \cite{blum},
\begin{equation}
\frac{d}{dt}\hat{\rho}=-\frac{i}{\hbar }\left[ \hat{H},\hat{\rho}\right] +%
\hat{L}(\hat{\rho}),  \label{meq1}
\end{equation}%
where $\hat{L}(\hat{\rho})$ is the relaxation operator. If there are $M$
states in a given basis $|\alpha \rangle $, Eq.~(\ref{meq1}) corresponds to $%
\frac{1}{2}M(M+1)$ equations for the matrix elements $\rho _{\alpha \beta
}=\rho _{\beta \alpha }^{\ast }$. The number of equations that need to be
solved can be reduced to $M$ within the method of the stochastic equation of
evolution for the state vector \cite{tokman2020,chen2021}. This becomes
possible if the structure of the relaxation operator permits representing
the right-hand side of Eq.~(\ref{meq1}) in the form
\begin{equation}
-\frac{i}{\hbar }\left[ \hat{H},\hat{\rho}\right] +\hat{L}(\hat{\rho})=-%
\frac{i}{\hbar }(\hat{H}_{eff}\hat{\rho}-\hat{\rho}\hat{H}_{eff}^{\dagger
})+\delta \hat{L}(\hat{\rho}),  \label{meq2}
\end{equation}%
where $\hat{H}_{eff}=\hat{H}+\hat{H}^{\left( ah\right) }$ is an effective
non-Hermitian Hamiltonian. In the simplest case of inelastic relaxation (i.e., ignoring the elastic scattering processes leading to pure dephasing)  in a
system with nonequidistant spectrum and zero temperature the anti-Hermitian
part $\hat{H}^{\left( ah\right) }$ of the effective Hamiltonian describes
the decay of the state populations and quantum coherences, whereas the term $%
\delta \hat{L}(\hat{\rho})$ describes an increase in the populations of
lower states due to relaxation from higher states \cite{fain,Scully1997}.

Within the Markovian models of relaxation the stochastic equation for the
state vector takes the form
\begin{equation}
\frac{d}{dt}\left\vert \Psi \right\rangle =-\frac{i}{\hbar }\hat{H}%
_{eff}\left\vert \Psi \right\rangle -\frac{i}{\hbar }\left\vert \mathfrak{R}%
\right\rangle .  \label{S eq}
\end{equation}%
In Eq.~(\ref{S eq}) the vector $\left\vert \mathfrak{R}\right\rangle $ is a
stochastic Langevin source with the following statistical properties:
\begin{equation}
\overline{\left\vert \mathfrak{R}\right\rangle }=0,\ \ \ \overline{\mathfrak{%
R}_{\alpha }\left( t^{\prime }\right) \mathfrak{R}_{\beta }^{\ast }\left(
t^{\prime \prime }\right) }=\hbar ^{2}\delta \left( t^{\prime }-t^{\prime
\prime }\right) D_{\alpha \beta },\ \ \ D_{\alpha \beta }=\left\langle
\alpha \right\vert \delta \hat{L}(\hat{\rho})\left\vert \beta \right\rangle
_{\hat{\rho}\Longrightarrow \overline{\left\vert \Psi \right\rangle
\left\langle \Psi \right\vert }},  \label{statistical properties}
\end{equation}
the overbar $\overline{\left( \cdots \right) }$ means averaging over the
noise statistics, $\mathfrak{R}_{\alpha }=\left\langle \alpha \right\vert
\left. \mathfrak{R}\right\rangle $ . The dyadics $\overline{C_{\alpha
}C_{\beta }^{\ast }}$ in Eqs.~(\ref{S eq}),(\ref{statistical properties})
where $C_{\alpha }=\left\langle \alpha \right\vert \left. \Psi \right\rangle
$ correspond to the density matrix elements $\rho _{\alpha \beta }$ in the
master equation (see the proof in \cite{tokman2020}).

The observables in the method of the stochastic equation are determined as
\begin{equation*}
g=\overline{\left\langle \Psi \right\vert \hat{g}\left\vert \Psi
\right\rangle },
\end{equation*}%
where $\hat{g}$ is an operator corresponding to the physical quantity $g$.
This definition differs from a standard one by an additional averaging over
the noise statistics. The choice of operators $\hat{H}^{\left( ah\right) }$
and correlators $D_{\alpha \beta }$ should ensure the conservation of the
norm of the stochastic vector, $\overline{\left\langle \Psi \right\vert
\left. \Psi \right\rangle }=1$, and bring the system to a physically
reasonable steady state in the absence of an external perturbation.

Note that the stochastic equation of the type of Eq.~(\ref{S eq}) is
conceptually similar to the method of quantum jumps which is often used in
numerical Monte-Carlo simulations \cite{Plenio1998,Scully1997}. The origin
of any stochastic equation approach as opposed to the density matrix
formalism is of course rooted in the Langevin formalism. An important
example of using the Langevin formalism in the derivation of the
fluctuation-dissipation theorem can be found in \cite{Landau1965}.

Another widely used method to include the effects of dissipation in quantum
optics is the Heisenberg-Langevin approach \cite{Scully1997,Gardiner2004}.
However, when applied to the dynamics of strongly coupled systems, the
Heisenberg equations become nonlinear (see, e.g., \cite{Scully1997}),
whereas the stochastic equation for the state vector, Eq.~(\ref{S eq}), is
always linear, which is an important advantage of this method.

The representation of the type shown in Eq.~(\ref{meq2}) is possible, in
particular, for the Lindblad relaxation operator. Here we will use the
Lindbladian $\hat{L}(\hat{\rho})$ in the case of independent dissipative
reservoirs for the field and the qubits and at zero temperature. The case of
arbitrary temperatures is considered in \cite{tokman2020}. Note that for a qubit with the
transition in the visible or near-IR range even a room-temperature reservoir
is effectively at zero temperature.
\begin{equation}
L(\hat{\rho})=-\Sigma _{j}[\frac{\gamma _{j}}{2}(\hat{\sigma}_{j}^{\dagger }%
\hat{\sigma}_{j}\hat{\rho}+\hat{\rho}\hat{\sigma}_{j}^{\dagger }\hat{\sigma}%
_{j}-2\hat{\sigma}_{j}\hat{\rho}\hat{\sigma}_{j}^{\dagger })]-\frac{\mu }{2}(%
\hat{c}^{\dagger }\hat{c}\hat{\rho}+\hat{\rho}\hat{c}^{\dagger }\hat{c}-2%
\hat{c}\hat{\rho}\hat{c}^{\dagger }),  \label{Lindbladian at 0 T}
\end{equation}%
which gives%
\begin{equation}
\hat{H}_{eff}=\hat{H}-i\hbar \frac{1}{2}\left( \sum_{j}\gamma _{j}\hat{\sigma%
}_{j}^{\dagger }\hat{\sigma}_{j}+\mu \hat{c}^{\dagger }\hat{c}\right) ,
\label{efh}
\end{equation}%
\begin{equation}
\delta \hat{L}(\hat{\rho})=\sum_{j}\gamma _{j}\hat{\sigma}_{j}\hat{\rho}\hat{%
\sigma}_{j}^{\dagger }+\mu \hat{c}\hat{\rho}\hat{c}^{\dagger }.
\label{delta LO}
\end{equation}%
Here the relaxation constants $\mu $ and $\gamma _{i}$ are determined by the
cavity Q-factor and inelastic relaxation of the qubits, respectively.
Elastic relaxation processes (pure dephasing) are included later in this section.

Consider state vectors $\Psi $ of the type given in Eq.~(\ref{Psi}).
Assuming different detunings $\Delta _{i}=\frac{W_{i}}{\hbar }-\omega $ for
each qubit in the most general case, we obtain a set of stochastic equations
\begin{equation}
\dot{C}_{00}+\left( i\frac{\omega }{2}+\gamma _{00}\right) C_{00}=-\frac{i}{%
\hbar }\mathfrak{R}_{00};  \label{C00 dotted}
\end{equation}%
\begin{equation}
\dot{C}_{10}+\left( i\frac{3\omega }{2}+\gamma _{10}\right)
C_{10}-i\sum_{j=1}^{N}\Omega _{Rj}^{\ast }C_{0j}=-\frac{i}{\hbar }\mathfrak{R%
}_{10},  \label{C10 dotted}
\end{equation}%
\begin{equation}
\dot{C}_{0j}+\left( i\frac{3\omega }{2}+i\Delta _{j}+\gamma _{0j}\right)
C_{0j}-i\Omega _{Rj}C_{10}=-\frac{i}{\hbar }\mathfrak{R}_{0j},
\label{C0j dotted}
\end{equation}%
where the relaxation constants are related to the EM field and qubit
relaxation constants in the Lindbladian Eq.~(\ref{Lindbladian at 0 T}) by
\begin{equation}
\gamma _{00}=0,\ \gamma _{10}=\frac{\mu }{2},\ \gamma _{0j}=\frac{\gamma _{j}%
}{2}.  \label{relaxation constants}
\end{equation}%
The noise properties are given by%
\begin{equation}
\overline{\mathfrak{R}_{\alpha n}^{\ast }\left( t^{\prime }\right) \mathfrak{%
R}_{\beta m}\left( t^{\prime \prime }\right) }=\hbar ^{2}\delta _{\alpha
\beta }\delta _{nm}D_{\alpha n,\alpha n}\delta \left( t^{\prime }-t^{\prime
\prime }\right) ,  \label{noise properties}
\end{equation}%
\begin{equation}
D_{00,00}=\sum_{j=1}^{N}\gamma _{j}\overline{\left\vert C_{0j}\right\vert
^{2}}+\mu \overline{\left\vert C_{10}\right\vert ^{2}},\ D_{10,10}=0,\
D_{0j,0j}=0.  \label{D_abcd}
\end{equation}%

To include elastic relaxation (pure dephasing) in the Lindbladian Eq.~(\ref{Lindbladian at 0
T}) we need to add the term \cite{fain}%
\begin{equation*}
L^{\left( el\right) }(\hat{\rho})=-\Sigma _{j}[\frac{\gamma _{j}^{\left(
el\right) }}{2}(\hat{\sigma}_{zj}\hat{\sigma}_{zj}^{\dagger }\hat{\rho}+\hat{%
\rho}\hat{\sigma}_{zj}\hat{\sigma}_{zj}^{\dagger }-2\hat{\sigma}%
_{zj}^{\dagger }\hat{\rho}\hat{\sigma}_{zj})],
\end{equation*}%
where $\hat{\sigma}_{zj}=\hat{\sigma}_{zj}^{\dagger }=\left\vert
1_{j}\right\rangle \left\langle 1_{j}\right\vert -\left\vert
0_{j}\right\rangle \left\langle 0_{j}\right\vert $ and $\gamma _{j}^{\left(
el\right) }$ is an elastic relaxation constant. Recent analysis \cite%
{tokman2020,chen2021} shows that for two-level qubits the elastic processes
can be included by making the following replacements in the expressions for $%
\gamma _{0j}$ and $D_{0j,0j}$: $\gamma _{0j}\Longrightarrow \gamma _{0j}+$ $%
\gamma _{j}^{\left( el\right) }$, $D_{0j,0j}\Longrightarrow
D_{0j,0j}+2\gamma _{j}^{\left( el\right) }\overline{\left\vert
C_{0j}\right\vert ^{2}}$. These relationships lead to standard relaxation
timescales of populations, $T_{1j}=\frac{1}{\gamma _{j}}$ and coherence, $%
T_{2j}=\frac{1}{\frac{1}{2T_{1j}}+\gamma _{j}^{\left( el\right) }}$ \cite%
{fain}. Therefore, including pure decoherence processes leads to
corrections in the last of Eqs.~(\ref{relaxation constants}) and the last of
Eqs.~(\ref{D_abcd}), namely%
\begin{equation}
\gamma _{0j}=\frac{\gamma _{j}}{2}+\gamma _{j}^{\left( el\right) },\
D_{0j,0j}=2\gamma _{j}^{\left( el\right) }\   \label{corrections due to edp}
\end{equation}

Taking into account Eqs.~(\ref{corrections due to edp}), it is easy
to show that for any set of elastic scattering rates Eqs.~(\ref{C00 dotted}%
)-(\ref{C0j dotted}) conserve the norm:
\begin{equation}
\sum_{j=1}^{N}\overline{\left\vert C_{0j}\right\vert ^{2}}+\overline{%
\left\vert C_{10}\right\vert ^{2}}+\overline{\left\vert C_{00}\right\vert
^{2}}=1,  \label{norm condition}
\end{equation}%
and Eqs.~(\ref{C10 dotted})-(\ref{C0j dotted}) preserve the following
relationship which includes only the rates of \textit{inelastic} relaxation:
\begin{equation}
\frac{d}{dt}\left( \sum_{j=1}^{N}\overline{\left\vert C_{0j}\right\vert ^{2}}%
+\overline{\left\vert C_{10}\right\vert ^{2}}\right) =-\sum_{j=1}^{N}\gamma
_{j}\overline{\left\vert C_{0j}\right\vert ^{2}}-\mu \overline{\left\vert
C_{10}\right\vert ^{2}}.  \label{relationship}
\end{equation}

In the absence of elastic scattering one can drop the noise sources $-\frac{i%
}{\hbar }\mathfrak{R}_{10}$ and $-\frac{i}{\hbar }\mathfrak{R}_{0j}$ in the
right hand side of Eqs.~(\ref{C10 dotted}) and (\ref{C0j dotted}). Including
these noise sources would lead to the appearance of the terms in the
expressions for $C_{10}$ and $C_{0j}$ which are linear with respect to
corresponding random functions. Their contribution will be zero after
averaging over the noise statistics, provided the autocorrelators $D_{10,10}$
and $D_{0j,0j}$, and all \textquotedblleft off-diagonal\textquotedblright\
correlators (see Eqs.~(\ref{noise properties}),(\ref{D_abcd})) are equal to
zero. Therefore, the equations for $C_{10}$ and $C_{0j}$ become simplified:%
\begin{equation}
\dot{C}_{10}+\left( i\frac{3\omega }{2}+\frac{\mu }{2}\right)
C_{10}-i\sum_{j=1}^{N}\Omega _{Rj}^{\ast }C_{0j}=0,  \label{c10 eq sp}
\end{equation}%
\begin{equation}
\dot{C}_{0j}+\left( i\frac{3\omega }{2}+i\Delta _{j}+\frac{\gamma _{j}}{2}%
\right) C_{0j}-i\Omega _{Rj}C_{10}=0.  \label{c0j eq sp}
\end{equation}%
The equation for $C_{00}$ still contains a corresponding noise source term,%
\begin{equation}
\dot{C}_{00}+i\frac{\omega }{2}C_{00}=-\frac{i}{\hbar }\mathfrak{R}_{00}.
\label{C00 TD}
\end{equation}

Eqs.~(\ref{c10 eq sp})-(\ref{C00 TD}) can be considered an improvement over
the Weisskopf-Wigner approximation because, unlike the latter, they conserve
the norm of the state vector; see Eq.~(\ref{norm condition}). We will use
Eqs.~(\ref{c10 eq sp})-(\ref{C00 TD}) for further analysis to avoid overly
cumbersome expressions without losing any significant physics.

\subsection{Two-qubit dynamics in the presence of dissipation}

In this case $j=1,2$. To maintain the same notations as in Sec.
III, we will seek the state vector $\Psi $ in the form of Eq.~(\ref{Psi}).
We will rewrite Eqs.~(\ref{c10 eq sp})-(\ref{C00 TD}) after adapting the
notations in Eq.~(\ref{Psi}) to the case of two qubits and one photon mode: $%
C_{0j=1}\Rightarrow C_{010}$, $C_{0j=2}\Rightarrow C_{001}$, $%
C_{10}\Rightarrow C_{100}$, $C_{00}\Rightarrow C_{000}$, $\mathfrak{R}%
_{00}\Rightarrow \mathfrak{R}_{000}$, which gives%
\begin{equation}
\dot{C}_{100}+\left( i\frac{3\omega }{2}+\frac{\mu }{2}\right)
C_{100}-i(\Omega _{R1}^{\ast }C_{010}+\Omega _{R2}^{\ast }C_{001})=0,
\label{C100 dotted}
\end{equation}
\begin{equation}
\dot{C}_{010}+(i\frac{3\omega }{2}+i\Delta _{1}+\frac{\gamma _{1}}{2}%
)C_{010}-i\Omega _{R1}C_{100}=0,  \label{C010 dotted}
\end{equation}
\begin{equation}
\dot{C}_{001}+(i\frac{3\omega }{2}+i\Delta _{2}+\frac{\gamma _{2}}{2}%
)C_{001}-i\Omega _{R2}C_{100}=0,  \label{C001 dotted}
\end{equation}%
\begin{equation}
\dot{C}_{000}+i\frac{\omega }{2}C_{000}=-\frac{i}{\hbar }\mathfrak{R}_{000},
\label{C000 dotted}
\end{equation}%
where%
\begin{equation}
\overline{\mathfrak{R}_{000}}=0,\ \ \ \overline{\mathfrak{R}_{000}^{\ast
}\left( t^{\prime }\right) \mathfrak{R}_{000}\left( t^{\prime \prime
}\right) }=\hbar ^{2}D_{000,000}\delta \left( t^{\prime }-t^{\prime \prime
}\right) ,\ \ \ D_{000,000}=\gamma _{1}\overline{\left\vert
C_{010}\right\vert ^{2}}+\gamma _{2}\overline{\left\vert C_{001}\right\vert
^{2}}+\mu \overline{\left\vert C_{100}\right\vert ^{2}}.  \label{SP}
\end{equation}%
Making standard substitutions in Eqs.~(\ref{C100 dotted})-(\ref{C001 dotted}%
),%
\begin{equation*}
\left(
\begin{array}{c}
C_{100} \\
C_{010} \\
C_{001}%
\end{array}%
\right) =\left(
\begin{array}{c}
G_{100} \\
G_{010} \\
G_{001}%
\end{array}%
\right) e^{-i\omega \frac{3}{2}t},\ \ G_{\cdots }\propto e^{\Gamma t},
\end{equation*}%
We obtain%
\begin{equation}
\left(
\begin{array}{ccc}
\Gamma +\frac{\mu }{2} & -i\Omega _{R1}^{\ast } & -i\Omega _{R2}^{\ast } \\
-i\Omega _{R1} & \Gamma +i\Delta _{1}+\frac{\gamma _{1}}{2} & 0 \\
-i\Omega _{R2} & 0 & \Gamma +i\Delta _{2}+\frac{\gamma _{2}}{2}%
\end{array}%
\right) \left(
\begin{array}{c}
G_{100} \\
G_{010} \\
G_{001}%
\end{array}%
\right) =0,  \label{eigen Eq}
\end{equation}%
which gives the dispersion equation for frequency eigenvalues:

\begin{equation}
(\Gamma +\frac{\mu }{2})(\Gamma +i\Delta _{1}+\frac{\gamma _{1}}{2})(\Gamma
+i\Delta _{2}+\frac{\gamma _{2}}{2})+\left\vert \Omega _{R1}\right\vert
^{2}(\Gamma +i\Delta _{2}+\frac{\gamma _{2}}{2})+\left\vert \Omega
_{R2}\right\vert ^{2}(\Gamma +i\Delta _{1}+\frac{\gamma _{1}}{2})=0.
\label{dispersion equation}
\end{equation}%
To avoid cumbersome algebra, we consider two important limiting cases.

%%%%%%%%%%%%%%%%%%%%%%%%%%%%%%%%%%%%%%%%%%

\subsubsection{Very different qubits}

This is the case when one qubit is close to resonance with a cavity
mode whereas another qubit is \textquotedblleft
off-resonant\textquotedblright :
\begin{equation}
\left\vert i\Delta _{2}+\frac{\gamma _{2}}{2}\right\vert \gg \left\vert
\Omega _{R2}\right\vert ,\ \left\vert i\Delta _{1}+\frac{\gamma _{1}}{2}%
\right\vert \leq \left\vert \Omega _{R1}\right\vert .\   \label{VDQC}
\end{equation}

 If inequalities~(\ref{VDQC}) are satisfied, an asymptotic solution
of Eq.~(\ref{dispersion equation}) has the following roots,%
\begin{equation}
\Gamma _{1}\approx -i\Delta _{2}-\frac{\gamma _{2}}{2},\ \ \Gamma
_{2,3}\approx -\frac{i\Delta _{1}+\frac{\gamma _{1}}{2}+\frac{\mu }{2}}{2}%
\pm i\sqrt{\left\vert \Omega _{R1}\right\vert ^{2}-\frac{\left( i\Delta _{1}+%
\frac{\gamma _{1}}{2}-\frac{\mu }{2}\right) ^{2}}{4}}  \label{roots}
\end{equation}%
up to the terms of the order of \ $\frac{\left\vert \Omega _{R2}\right\vert
^{2}}{\left\vert i\Delta _{2}+\frac{\gamma _{2}}{2}\right\vert ^{2}}$ and $%
\frac{\left\vert \Omega _{R2}\right\vert ^{2}}{\left\vert \Omega
_{R1}\right\vert \left\vert i\Delta _{2}+\frac{\gamma _{2}}{2}\right\vert }$
respectively. For the matrix in the left-hand side of Eq.~(\ref{eigen Eq})
the eigenvalue $\Gamma _{1}$ corresponds to the eigenstate $\left(
\begin{array}{c}
G_{100} \\
G_{010} \\
G_{001}%
\end{array}%
\right) =\left(
\begin{array}{c}
0 \\
0 \\
1%
\end{array}%
\right) $ , whereas eigenvalues $\Gamma _{2,3}$ correspond to the
eigenstates $\left(
\begin{array}{c}
G_{100} \\
G_{010} \\
G_{001}%
\end{array}%
\right) \propto \left(
\begin{array}{c}
1 \\
\frac{i\Omega _{R1}}{\Gamma _{2,3}+i\Delta _{1}+\frac{\gamma _{1}}{2}} \\
0%
\end{array}%
\right) $ . Similarly to Sec. III, the system is split into the nonresonant
qubit and a coupled system ``cavity mode + resonant
qubit''. In the presence of dissipation the conditions for
such decoupling in Eq.~(\ref{VDQC}) depend on the relaxation constants and
can be satisfied even at resonance $\Delta _{1}=\Delta _{2}=0$ if one of the
qubits decays fast enough: $\gamma _{2}\gg \gamma _{1}$, $\mu $, $\left\vert
\Omega _{R2}\right\vert $.

\subsubsection{Identical qubits}

Now we can put $\Delta _{1}=\Delta _{2}=\Delta $ and\ $\gamma
_{1}=\gamma _{2}=\gamma $ in Eqs.~(\ref{C100 dotted})-(\ref{C001 dotted});
the values of \ $\Omega _{R1}$ and $\Omega _{R2}$ are still different,
because the qubits can be located inside the cavity in places with different
amplitudes of the field mode; see Eqs.~(\ref{E}) and (\ref{nc-const}).

The eigenvalue equation~(\ref{dispersion equation}) takes the form
\begin{equation}
(\Gamma +i\Delta +\frac{\gamma }{2})\left[ (\Gamma +\frac{\mu }{2})(\Gamma
+i\Delta +\frac{\gamma }{2})+\left\vert \Omega _{R1}\right\vert
^{2}+\left\vert \Omega _{R2}\right\vert ^{2}\right] =0,  \label{DE}
\end{equation}%
which has the solutions
\begin{equation}
\Gamma _{1}=-i\Delta -\frac{\gamma }{2},\ \ \Gamma _{2,3}=-\frac{i\Delta +%
\frac{\gamma }{2}+\frac{\mu }{2}}{2}\pm i\sqrt{\left\vert \Omega
_{R1}\right\vert ^{2}+\left\vert \Omega _{R2}\right\vert ^{2}-\frac{\left(
i\Delta +\frac{\gamma }{2}-\frac{\mu }{2}\right) ^{2}}{4}}  \label{Roots}
\end{equation}%
One can see that $\mathrm{Re}[\Gamma _{2,3}]<0$ is always satisfied. The
general solution has the form
\begin{equation}
\left(
\begin{array}{c}
C_{100} \\
C_{010} \\
C_{001}%
\end{array}%
\right) =e^{-i\frac{3\omega }{2}t}\times \left[ \frac{Ae^{\Gamma _{1}t}}{%
\sqrt{N_{1}}}\left(
\begin{array}{c}
0 \\
1 \\
-\frac{\Omega _{R1}^{\ast }}{\Omega _{R2}^{\ast }}%
\end{array}%
\right) +\frac{Be^{\Gamma _{2}t}}{\sqrt{N_{2}}}\left(
\begin{array}{c}
1 \\
\frac{i\Omega _{R1}}{\Gamma _{2}+i\Delta +\frac{\gamma }{2}} \\
\frac{i\Omega _{R2}}{\Gamma _{2}+i\Delta +\frac{\gamma }{2}}%
\end{array}%
\right) +\frac{Ce^{\Gamma _{3}t}}{\sqrt{N_{3}}}\left(
\begin{array}{c}
1 \\
\frac{i\Omega _{R1}}{\Gamma _{3}+i\Delta +\frac{\gamma }{2}} \\
\frac{i\Omega _{R2}}{\Gamma _{3}+i\Delta +\frac{\gamma }{2}}%
\end{array}%
\right) \right] ,  \label{eigenstates}
\end{equation}%
where $N_{1}=1+\frac{\left\vert \Omega _{R1}\right\vert ^{2}}{\left\vert
\Omega _{R2}\right\vert ^{2}}$,\ $N_{2,3}=1+\frac{\left\vert \Omega
_{R1}\right\vert ^{2}}{\left\vert \Gamma _{2,3}+i\Delta +\frac{\gamma }{2}%
\right\vert ^{2}}+\frac{\left\vert \Omega _{R2}\right\vert ^{2}}{\left\vert
\Gamma _{2,3}+i\Delta +\frac{\gamma }{2}\right\vert ^{2}}$; the constants $A$%
, $B$ and $C$ are determined by initial conditions.

The first term in the sum in Eq.~(\ref{eigenstates}) corresponds to the
solution in which the qubits are decoupled from the field as a result of
destructive interference of their quantum states.

This interference leads to interesting consequences in the case typical for
plasmonic nanocavities, when the decay of the cavity mode is much faster
than the relaxation of the qubits, i.e. $\mu \gg \gamma $. At short enough
time intervals $\frac{\gamma }{2}t\ll 1$, one can put $\gamma =0$ in both
Eqs.~(\ref{Roots}). In this case the eigenfrequencies are%
\begin{equation*}
\Gamma _{1}=-i\Delta ,\ \ \Gamma _{2,3}=-\frac{\mu }{4}-i\left( \frac{\Delta
}{2}\mp \sqrt{\left\vert \Omega _{R1}\right\vert ^{2}+\left\vert \Omega
_{R2}\right\vert ^{2}-\frac{\left( i\Delta -\frac{\mu }{2}\right) ^{2}}{4}}%
\right)
\end{equation*}%
Therefore, the first term in the brackets in Eq.~(\ref{eigenstates}) does
not decay whereas other terms decay fast, with the rate $\sim \frac{\mu }{4}$%
. It means that at times $\frac{\mu }{4}t\gg 1$ the dissipative dynamics
drives the system to the asymptotic steady state in which the qubits are in
an entangled state completely decoupled from the cavity mode, whereas the EM
field is in its vacuum state:
\begin{equation}
\Psi =e^{-i\frac{\omega }{2}t}\left\vert 0\right\rangle \left[ e^{-i\frac{W}{%
\hbar }t}A\frac{\left\vert 0\right\rangle \left\vert 1\right\rangle
\left\vert 0\right\rangle -\frac{\Omega _{R1}^{\ast }}{\Omega _{R2}^{\ast }}%
\left\vert 0\right\rangle \left\vert 0\right\rangle \left\vert
1\right\rangle }{\sqrt{N_{1}}}+C_{000}\left\vert 0\right\rangle \left\vert
0\right\rangle \left\vert 0\right\rangle \right] ,  \label{quantum state}
\end{equation}%
The ground-state amplitude averaged over the noise statistics satisfies $%
\overline{C_{000}}=C_{000}\left( 0\right) $, $\overline{\left\vert
C_{000}\right\vert ^{2}}=1-\left\vert A\right\vert ^{2}$. It is obvious from
the latter equality and the conservation of the norm that $\left\vert
A\right\vert ^{2}$ determines the steady-state value of the average qubit
population at times $\frac{1}{\mu }\ll t\ll \frac{1}{\gamma }$. The
expression for $\left\vert A\right\vert ^{2}$ is
\begin{equation}
|A|^{2}=\frac{1-|C_{000}(0)|^{2}-|C_{100}(0)|^{2}}{1+|Z|^{2}}\left( 1+\frac{%
\left\vert \Omega _{R1}\right\vert ^{2}}{\left\vert \Omega _{R2}\right\vert
^{2}}\right) \left[ \frac{\left\vert \frac{|\Omega _{R2}|^{2}}{|\Omega
_{R1}|^{2}}-Z\frac{\Omega _{R2}^{\ast }}{\Omega _{R1}^{\ast }}\right\vert
^{2}}{\left( 1+\frac{\left\vert \Omega _{R2}\right\vert ^{2}}{\left\vert
\Omega _{R1}\right\vert ^{2}}\right) ^{2}}\right] ,  \label{A squared}
\end{equation}%
where $Z=\frac{C_{001}(0)}{C_{010}(0)}$.

The value of $|A|^{2}$ reaches a maximum when $\mathrm{Arg}[Z]=\pi -\mathrm{%
Arg}\left[ \frac{\Omega _{R2}^{\ast }}{\Omega _{R1}^{\ast }}\right] $ and $%
|Z|=\left\vert \frac{\Omega _{R1}}{\Omega _{R2}}\right\vert $, which
corresponds to $C_{001}(0)=-C_{010}(0)\frac{\Omega _{R1}^{\ast }}{\Omega
_{R2}^{\ast }}$ and
\begin{equation}
|A|^{2}=1-|C_{000}(0)|^{2}-|C_{100}(0)|^{2}.  \label{a2}
\end{equation}%
Clearly, at its maximum the total qubit occupation is equal to its initial
value, i.e. it remains constant. For any other initial conditions, the
initial qubit population dissipates partially, until it reaches the
decoupled entangled steady state with a smaller value of $|A|^{2}$.

%%%%%%%%%%%%%%%%%%%%%%%%%%%%%%%%%%%%%%%%

\begin{figure}[htb]
\includegraphics[width=0.5\textwidth]{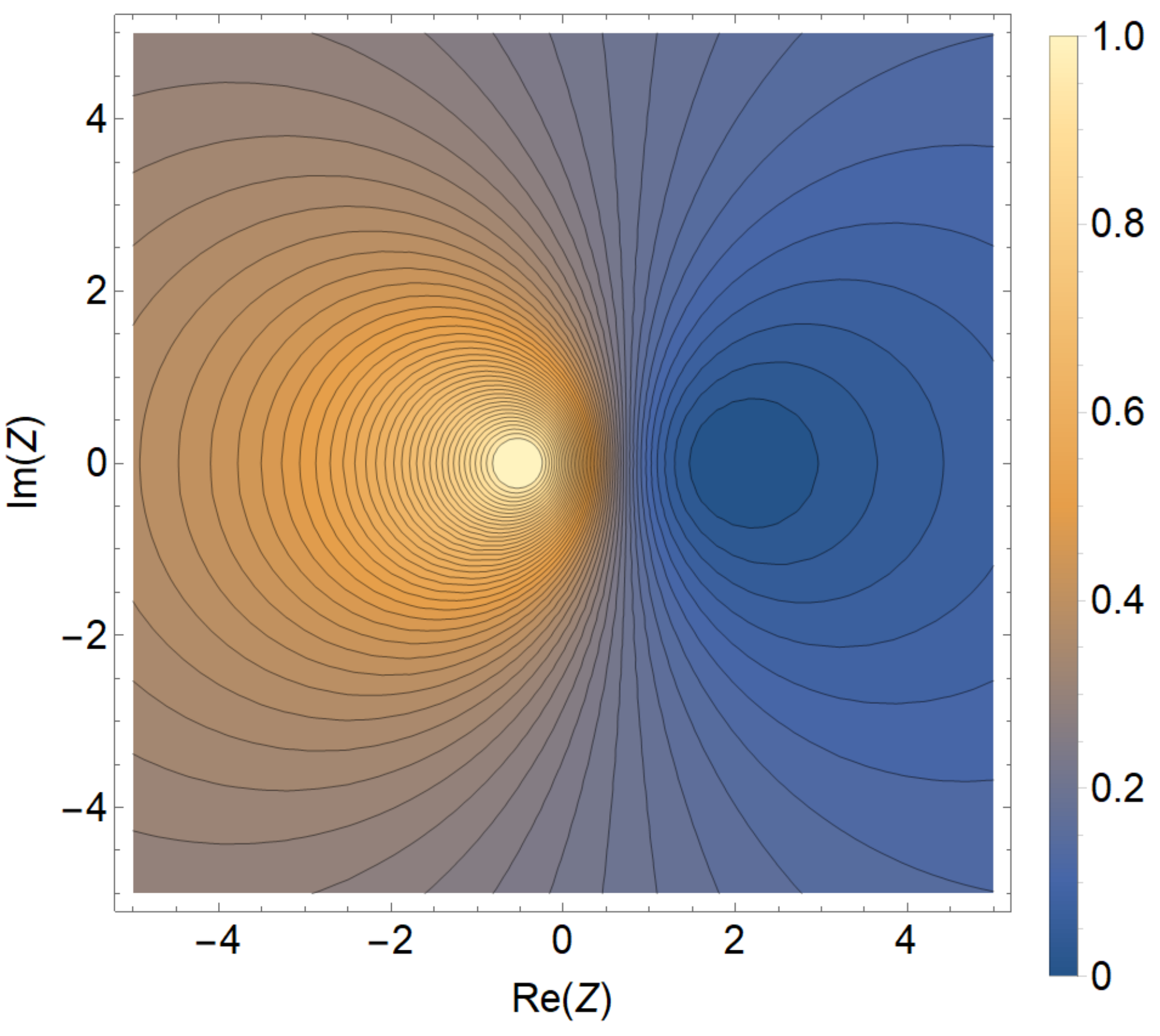}
\caption{ The contour plot of the value of $|A|^2$ on the complex $Z$ plane
for $\Omega_{R2} = 2 \Omega_{R1}$ and $C_{000}(0)=C_{100}(0) = 0$.}
\label{asquared}
\end{figure}

This is illustrated in Fig.~\ref{asquared} which shows the contour plot of
the value of $|A|^2$ on the complex $Z$ plane for real $\Omega_{R2} = 2
\Omega_{R1}$ and a particular set of initial conditions $%
C_{000}(0)=C_{100}(0) = 0$. For this particular choice of parameters, the
maximum of $|A|^2$ is reached on the real axis at the point $Z = -\frac{1}{%
\sqrt{2}}$ as follows from the above and is shown in the figure.

A conceptually similar effect of dissipation-driven decoupling of the
dynamical system from the dissipative reservoir as a result of destructive
interference was found in \cite{chen2021} for one qubit coupled to two EM
modes in a cavity with a time-modulated parameter (modulation-induced
transparency).

One can also identify the initial conditions for which a \textit{constructive} interference of qubit
quantum states occurs, which leads to a rapid complete energy transfer from the
qubits to the cavity field followed by its decay. This set of initial conditions corresponds to the minimum value $|A|^2 = 0$ in Fig.~\ref{asquared}. In this case the excited qubit population decays completely to the ground state due to coupling to the leaky cavity mode before the relaxation in the qubits even kicks in. 

\subsection{The formation of a decoupled entangled state of
N qubits}

Consider again a system of N identical qubits. We will again use Eqs.~(\ref%
{c10 eq sp})-(\ref{C00 TD}) with $\Delta _{j}=\gamma _{j}=0$, $\Omega
_{Rj}=\Omega _{R}$:%
\begin{equation}
\dot{C}_{10}+\left( i\frac{3\omega }{2}+\frac{\mu }{2}\right) C_{10}-i\Omega
_{R}^{\ast }\sum_{j=1}^{N}C_{0j}=0,  \label{c10 dotted}
\end{equation}%
\begin{equation}
\dot{C}_{0j}+i\frac{3\omega }{2}C_{0j}-i\Omega _{R}C_{10}=0.
\label{c0j dotted}
\end{equation}%
After making the same substitutions as in Sec. II,%
\begin{equation*}
\sum_{j=1}^{N}C_{0j}=F,\ \ \left(
\begin{array}{c}
C_{10} \\
F%
\end{array}%
\right) =\left(
\begin{array}{c}
G_{10} \\
G_{F}%
\end{array}%
\right) e^{-i\frac{3\omega }{2}t},\ \ \left(
\begin{array}{c}
G_{10} \\
G_{F}%
\end{array}%
\right) \propto e^{\Gamma t},
\end{equation*}%
we obtain%
\begin{equation}
\left(
\begin{array}{cc}
\Gamma +\frac{\mu }{2} & -i\Omega _{R}^{\ast } \\
-iN\Omega _{R} & \Gamma%
\end{array}%
\right) \left(
\begin{array}{c}
G_{10} \\
G_{F}%
\end{array}%
\right) =0,  \label{Eigen Eq}
\end{equation}%
where%
\begin{equation}
\Gamma ^{2}+\frac{\mu }{2}\Gamma +N\left\vert \Omega _{R}\right\vert ^{2}=0.
\label{Dis Eq}
\end{equation}%
The resulting solution for the amplitudes $C_{10}$ and $F$ is%
\begin{equation}
\left(
\begin{array}{c}
C_{10} \\
F%
\end{array}%
\right) =e^{-\left( i\frac{3\omega }{2}+\frac{\mu }{4}\right) t}\left[ \frac{%
Ae^{i\Xi t}}{\sqrt{1+N}}\left(
\begin{array}{c}
1 \\
\frac{\Gamma _{1}+\frac{\mu }{2}}{i\Omega _{R}^{\ast }}%
\end{array}%
\right) +\frac{Be^{-i\Xi t}}{\sqrt{1+N}}\left(
\begin{array}{c}
1 \\
\frac{\Gamma _{2}+\frac{\mu }{2}}{i\Omega _{R}^{\ast }}%
\end{array}%
\right) \right] ,  \label{eigen vector}
\end{equation}%
where the constants $A$ and $B$ are determined by initial conditions,%
\begin{equation}
\Gamma _{1,2}=-\frac{\mu }{4}\pm i\Xi ,\ \Xi =\sqrt{N\left\vert \Omega
_{R}\right\vert ^{2}-\frac{\mu ^{2}}{16}}.  \label{Solutions}
\end{equation}%
The expression for $C_{10}\left( t\right) $ which follows from Eq.~(\ref%
{eigen vector}) can be substituted into Eq.~(\ref{c0j dotted}) in order to
determine the solution for the amplitudes $C_{0j}$:%
\begin{equation}
C_{0j}\left( t\right) =C_{0j}\left( 0\right) e^{-i\frac{3\omega }{2}%
t}+i\Omega _{R}e^{-i\frac{3\omega }{2}t}\int_{0}^{t}e^{i\frac{3\omega }{2}%
\tau }C_{10}\left( \tau \right) d\tau .  \label{C0j(t)}
\end{equation}%
The expression for $C_{00}\left( t\right) $ can be obtained using Eq.~(\ref%
{C00 TD}),%
\begin{equation}
C_{00}\left( t\right) =C_{00}\left( 0\right) e^{-i\frac{\omega }{2}t}+\delta
C_{00},  \label{C00(t)}
\end{equation}%
where%
\begin{equation}
\delta C_{00}=-\frac{i}{\hbar }e^{-i\frac{\omega }{2}t}\int_{0}^{t}e^{i\frac{%
\omega }{2}t^{\prime }}\mathfrak{R}_{00}\left( t^{\prime }\right) dt^{\prime
}.  \label{delta C00}
\end{equation}%
Since $\overline{\delta C_{00}}=0$, we obtain $\overline{\left\vert
C_{00}\left( t\right) \right\vert ^{2}}=\overline{\left\vert C_{00}\left(
0\right) \right\vert ^{2}}+\overline{\left\vert \delta C_{00}\right\vert ^{2}%
}$. The easiest way to calculate the averaged probability $\overline{%
\left\vert C_{00}\left( t\right) \right\vert ^{2}}$ is to use the
conservation of the norm of the state vector, Eq.~(\ref{norm condition}).

At times longer than the cavity decay time, $\mu t\rightarrow \infty $, we
always have $C_{10}=0$ and $F=\sum_{j=1}^{N}C_{0j}=0$. However, this does
not necessarily mean that $\sum_{j=1}^{N}\left\vert C_{0j}\right\vert ^{2}=0$%
. In other words, an ensemble of qubits may approach a quasi-steady-state
with a finite average energy which is decoupled from a dissipative cavity
field. This situation is similar to the destructive interference in a system
of two qubits considered in previous subsection IVB. The exact form of the
asymptotic final state depends on the initial conditions. Here we consider
two natural initial conditions which lead to very different evolutions of
the quantum state.

%%%%%%%%%%%%%%%%%%%%

\subsubsection{Only the cavity mode is excited initially}

\begin{equation}
\Psi (0)=\left\vert 1\right\rangle \Pi _{j=1}^{N}\left\vert
0_{j}\right\rangle.  \label{IS}
\end{equation}%
In this case Eq.~(\ref{eigen vector}) gives%
\begin{equation}
C_{10}=e^{-\left( i\frac{3\omega }{2}+\frac{\mu }{4}\right) t}\left[ \cos
\left( \Xi t\right) -\frac{\mu }{4\Xi }\sin \left( \Xi t\right) \right]
\label{C10}
\end{equation}%
and%
\begin{equation}
F=e^{-\left( i\frac{3\omega }{2}+\frac{\mu }{4}\right) t}i\frac{\Omega _{R}}{%
\Xi }\sin \left( \Xi t\right) .  \label{F}
\end{equation}%
Substituting Eq.~(\ref{C10}) into Eq.~(\ref{C0j(t)}) we arrive at%
\begin{eqnarray}
\Psi &=&e^{-\left( i\frac{3\omega }{2}+\frac{\mu }{4}\right) t}\left\{ \left[
\cos \left( \Xi t\right) -\frac{\mu }{4\Xi }\sin \left( \Xi t\right) \right]
\left\vert 1\right\rangle \Pi _{j=1}^{N}\left\vert 0_{j}\right\rangle +i%
\frac{\Omega _{R}}{\Xi }\sin \left( \Xi t\right) \left\vert 0\right\rangle
\sum_{j=1}^{N}\left\vert 1_{j}\right\rangle \Pi _{m\neq j}^{N}\left\vert
0_{m}\right\rangle \right\}  \notag \\
&&+e^{-i\frac{\omega }{2}t}C_{00}\left\vert 0\right\rangle \Pi
_{j=1}^{N}\left\vert 0_{j}\right\rangle  \label{WF}
\end{eqnarray}%
where%
\begin{equation}
\overline{C_{00}}=0,\ \overline{\ \left\vert C_{00}\left( 0\right)
\right\vert ^{2}}=1-e^{-\frac{\mu }{2}t}\left[ 1+\frac{\mu ^{2}}{8\Xi ^{2}}%
\sin ^{2}\left( \Xi t\right) -\frac{\mu }{2\Xi }\cos \left( \Xi t\right)
\sin \left( \Xi t\right) \right] .  \label{Cond}
\end{equation}%
The final state at long enough times $\mu t\rightarrow \infty $ is
factorized,%
\begin{equation}
\Psi \left( t\rightarrow \infty \right) =e^{-i\frac{\omega }{2}t}\left\vert
0\right\rangle \Pi _{j=1}^{N}\left\vert 0_{j}\right\rangle .  \label{FS}
\end{equation}%
This solution corresponds to zero probability of the excitation transfer
from the cavity field to an ensemble of qubits.

%%%%%%%%%%%%%%%%%%%%%%%%%%%

\subsubsection{The initial condition corresponds to the 
excitation of qubits whereas the cavity field is in the vacuum state}

The most general initial state of this kind can be written as
\begin{equation}
\Psi (0)=\sum_{j=1}^{N}C_{0j}\left( 0\right) \left\vert 0\right\rangle
\left\vert 1_{j}\right\rangle \Pi _{q\neq j}^{N}\left\vert
0_{q}\right\rangle .  \label{Initial state}
\end{equation}%
In this case Eq.~(\ref{eigen vector}) gives%
\begin{equation}
C_{10}=e^{-\left( i\frac{3\omega }{2}+\frac{\mu }{4}\right) t}iF\left(
0\right) \frac{\Omega _{R}^{\ast }}{\Xi }\sin \left( \Xi t\right)
\label{c10}
\end{equation}%
and%
\begin{equation}
F=e^{-\left( i\frac{3\omega }{2}+\frac{\mu }{4}\right) t}F\left( 0\right) %
\left[ \cos \left( \Xi t\right) +\frac{\mu }{4\Xi }\sin \left( \Xi t\right) %
\right] .  \label{f}
\end{equation}%
where $F\left( 0\right) =\sum_{j=1}^{N}C_{0j}\left( 0\right) $. Substituting
Eq.~(\ref{c10}) into Eq.~(\ref{C0j(t)}) and using the formula $%
\int_{0}^{\infty }e^{-\frac{\mu }{4}\tau }\sin \left( \Xi \tau \right) d\tau
=\frac{\Xi }{N\left\vert \Omega _{R}\right\vert ^{2}}$ for integration, we
obtain the solution for the amplitudes $C_{0j}$ at the times $\mu
t\rightarrow \infty $, but before the qubit decoherence kicks in, i.e. $%
\gamma t \ll 1$,
\begin{equation}
C_{0j}\left( t\rightarrow \infty \right) =\left[ C_{0j}\left( 0\right) -%
\frac{1}{N}\sum_{j=1}^{N}C_{0j}\left( 0\right) \right] e^{-i\frac{3\omega }{2%
}t}.  \label{C0j final}
\end{equation}%
It corresponds to the vacuum state of the field and the following entangled
state of the qubits,
\begin{eqnarray}
\Psi \left( t\rightarrow \infty \right) &=&e^{-i\frac{3\omega }{2}%
t}\sum_{j=1}^{N}\left[ C_{0j}\left( 0\right) \left( 1-\frac{1}{N}\right) -%
\frac{1}{N}\sum_{q\neq j}^{N}C_{0q}\left( 0\right) \right] \left\vert
0\right\rangle \left\vert 1_{j}\right\rangle \Pi _{q\neq j}^{N}\left\vert
0_{q}\right\rangle  \notag \\
&&+e^{-i\frac{\omega }{2}t}C_{00}\left\vert 0\right\rangle \Pi
_{j=1}^{N}\left\vert 0_{j}\right\rangle  \label{AW}
\end{eqnarray}%
where $\overline{C_{00}}=0$. The magnitude of $\overline{\left\vert
C_{00}\left( 0\right) \right\vert ^{2}}$ is easiest to find via the
conservation of the norm of the state vector, Eq.~(\ref{norm condition}): $%
\overline{\left\vert C_{00}\left( 0\right) \right\vert ^{2}}%
=1-\sum_{j=1}^{N}\left\vert C_{0j}\left( 0\right) \right\vert ^{2}$ , where
the values of $C_{0j}$ are given by Eq.~(\ref{C0j final}). \

As an example, consider an initial state in which only one qubit is excited.
Although no knowledge is needed regarding which one of the qubits is excited
(see the comment in Sec. II), for definiteness we will take the one with $%
j=1 $, i.e., $C_{01}\left( 0\right) =1$, $C_{0j\neq 1}\left( 0\right) =0$.
In this case $C_{01}\left( t\rightarrow \infty \right) =1-\frac{1}{N}$, $%
C_{0j\neq 1}\left( t\rightarrow \infty \right) =-\frac{1}{N}$, $\overline{%
\left\vert C_{00}\left( 0\right) \right\vert ^{2}}=\frac{1}{N}$. The state
vector in Eq.~(\ref{AW}) allows one to calculate the probabilities of three
possible events:

a) Dissipation of an excited qubit due to photon emission into the decaying
cavity field; the probability is $P_{rad}=\overline{\left\vert C_{00}\left(
0\right) \right\vert ^{2}}=\frac{1}{N}$.

b) Energy transfer from the excited qubit to other qubits through their
collective coupling to the EM cavity field; the probability is $%
P_{tr}=\left( N-1\right) \left\vert C_{0j\neq 1}\left( t\rightarrow \infty
\right) \right\vert ^{2}=\frac{N-1}{N^{2}}$.

c) An initially excited qubit preserves its initial state; the probability
is $P_{pr}=\left\vert C_{01}\left( t\rightarrow \infty \right) \right\vert
^{2}=\left( 1-\frac{1}{N}\right) ^{2}$.

To summarize, if there is only one qubit in the cavity, $N=1$, then its
excitation energy will be transferred to a resonant photon with probability $%
P_{rad}=1$ and quickly dissipate. For a relatively small number of qubits $%
N\geq 2$ (but with only one qubit initially excited), there are comparable
probabilities of all three events: $P_{rad}<1$, $P_{pr}\neq 0$ and $%
P_{tr}\neq 0$. If $N\gg 1$ we obtain $P_{pr}\gg P_{rad}\approx $ $%
P_{tr}\approx \frac{1}{N}$, i.e. the probability for an initially excited
qubit to preserve its state is much higher than the probabilities of energy
outcoupling to the field or other qubits.

Therefore, an ensemble of ground-state qubits effectively shields an excited
qubit from its outcoupling to a resonant cavity field. The shielding is due
to destructive interference of quantum states in an entangled state.

The same kind of screening exists for an arbitrary initial state of a
relatively small group of $J$ qubits if $J\ll N$:%
\begin{equation}
\Psi (0)=\left\vert 0\right\rangle \sum_{j=1}^{J}C_{0j}\left( 0\right)
\left\vert 0\right\rangle \left\vert 1_{j}\right\rangle \Pi _{q\neq
j}^{N}\left\vert 0_{q}\right\rangle ,  \label{ini state}
\end{equation}%
where $\sum_{j=1}^{J}\left\vert C_{0j}\left( 0\right) \right\vert ^{2}=1$.
Indeed, in this case Eq.~(\ref{AW}) has $C_{0;1\leq j\leq J}\left( 0\right)
\sim \frac{1}{\sqrt{J}}$, $\frac{1}{N}\sum_{q\neq j}^{N}C_{0j}\left(
0\right) =\frac{1}{N}\sum_{q\neq j}^{J}C_{0j}\left( 0\right) \sim \frac{%
\sqrt{J}}{N}$, i.e., the perturbation of the initial state is small: of the
order of $\frac{J}{N}$,%
\begin{equation*}
C_{0;1\leq j\leq J}\left( t\rightarrow \infty \right) =C_{0;1\leq j\leq
J}\left( 0\right) +o\left( \frac{J}{N}\right) .
\end{equation*}%
Similarly to the case of two qubits, one can also find an example of \textit{constructive} interference of qubit
quantum states which leads to a rapid complete energy transfer from the
qubits to the cavity field even if $N\gg 1$. This happens for the initial
state of the type
\begin{equation}
\Psi (0)=\frac{1}{\sqrt{N}}\sum_{j=1}^{N}\left\vert 0\right\rangle
\left\vert 1_{j}\right\rangle \Pi _{q\neq j}^{N}\left\vert
0_{q}\right\rangle ,  \label{ini wf}
\end{equation}%
Starting from this initial condition, the state vector approaches $\Psi
\left( t\rightarrow \infty \right) =\left\vert 0\right\rangle \Pi
_{j}^{N}\left\vert 0_{j}\right\rangle $ on the timescale $\frac{\mu }{4}t\gg
1$. Note that the initial state (\ref{ini wf}) which ensures constructive
interference for qubit radiation into the cavity mode corresponds to the
solution for the state vector in Eq.~(\ref{WF}) at the moments of time $\Xi
t=\frac{\pi }{2}+M\pi $ if we neglect dissipation, $\mu =0$.

%%%%%%%%%%%%%%%%%%%%%%%%%%%%%%%%%%%%%%%%%%%%%%%

\section{Emission from an ensemble of N qubits in a leaky cavity}

Detecting radiation from quantum emitters placed in nanocavities is one of
the most straightforward ways to study their quantum dynamics \cite%
{Scully1997, madsen2013, lodahl2015, chikkaraddy2016, pelton2018, park2019}.
The power spectrum received by the detector can be calculated as \cite%
{madsen2013,Scully1997}%
\begin{equation*}
P\left( \nu \right) =A\cdot S\left( \upsilon \right) ,
\end{equation*}%
where%
\begin{equation}
S\left( \upsilon \right) =\frac{1}{\pi }\mathrm{Re}\int_{0}^{\infty }d\tau
e^{i\upsilon \tau }\int_{0}^{\infty }dtK\left( t,\tau \right) ,  \label{S(v)}
\end{equation}%
\begin{equation}
K=\left\langle \Psi \left( 0\right) \right\vert \hat{c}^{\dagger }\left(
t\right) \hat{c}\left( t+\tau \right) \left\vert \Psi \left( 0\right)
\right\rangle ,  \label{K}
\end{equation}%
$\hat{c}^{\dagger }\left( t\right) $ and $\hat{c}\left( t\right) $ are
Heisenberg creation and annihilation operators for the cavity field, $\Psi
\left( 0\right) $ is an initial state of the system. The coefficient $A$ is
determined by the cavity design, spatial structure of the cavity field, and
detector properties.

These equations indicate that to calculate the power spectrum one has to
solve the Heisenberg-Langevin equations for the operators $\hat{c}\left(
t\right) $ and $\hat{c}^{\dagger }\left( t\right) $ \cite{Scully1997} and
evaluate the correlator including averaging over the noise statistics, $%
K\Rightarrow $ $\overline{\left\langle \Psi \left( 0\right) \right\vert \hat{%
c}^{\dagger }\left( t\right) \hat{c}\left( t+\tau \right) \left\vert \Psi
\left( 0\right) \right\rangle }$. However, the Heisenberg-Langevin equations
are nonlinear in the strong-coupling Rabi oscillations regime for a
single-photon field. Therefore, it is more convenient to utilize the
solution of the linear stochastic equation~(\ref{S eq}) for the state vector.

%%%%%%%%%%%%%%%%%%%%%%%%%%%%%%%%%%%%%%%%%%%%%%%

\subsection{Calculating the emission spectrum with the stochastic equation
for the state vector}

It is instructive to consider first the dissipationless problem. If $\hat{U}%
\left( t\right) $ is the evolution operator of the system, then the
correlator can be expressed as%
\begin{equation}
K=\left\langle \Psi \left( 0\right) \right\vert \hat{U}^{\dagger }\left(
t\right) \hat{c}^{\dagger }\hat{U}\left( t\right) \hat{U}^{\dagger }\left(
t+\tau \right) \hat{c}\hat{U}\left( t+\tau \right) \left\vert \Psi \left(
0\right) \right\rangle =\left\langle \hat{c}\Psi \left( t\right) \right\vert
\hat{U}\left( t\right) \hat{U}^{\dagger }\left( t+\tau \right) \left\vert
\hat{c}\Psi \left( t+\tau \right) \right\rangle ,  \label{k}
\end{equation}%
where $\hat{c}$ and $\hat{c}^{\dagger }$ are Schr\"{o}dinger's (constant)
operators. We will use the notation $\hat{U}\left( t\right) \Longrightarrow
\hat{U}_{t_{0}}\left( t^{\prime }\right) $ for the evolution operator which
indicates the initial moment of time $t_{0}$ and the duration of evolution $%
t^{\prime }=t-t_{0}$. With this notation we substitute $\hat{U}\left(
t\right) \Longrightarrow \hat{U}_{0}\left( t\right) $, $\hat{U}\left( t+\tau
\right) \Longrightarrow \hat{U}_{0}\left( t+\tau \right) =\hat{U}_{t}\left(
\tau \right) \hat{U}_{0}\left( t\right) $ in Eq.~(\ref{k}),%
\begin{equation*}
K=\left\langle \hat{c}\Psi \left( t\right) \right\vert \hat{U}_{0}\left(
t\right) \left( \hat{U}_{t}\left( \tau \right) \hat{U}_{0}\left( t\right)
\right) ^{\dagger }\left\vert \hat{c}\Psi \left( t+\tau \right)
\right\rangle =\left\langle \hat{c}\Psi \left( t\right) \right\vert \hat{U}%
_{0}\left( t\right) \hat{U}_{0}^{\dagger }\left( t\right) \hat{U}%
_{t}^{\dagger }\left( \tau \right) \left\vert \hat{c}\Psi \left( t+\tau
\right) \right\rangle .
\end{equation*}%
Taking into account the unitarity condition $\hat{U}_{0}\left( t\right) \hat{%
U}_{0}^{\dagger }\left( t\right) =1$, we obtain%
\begin{equation}
K=\left\langle \hat{U}_{t}\left( \tau \right) \hat{c}\Psi \left( t\right)
\right. \left\vert \hat{c}\Psi \left( t+\tau \right) \right\rangle .
\label{cor}
\end{equation}%
After introducing the notations%
\begin{equation}
\Psi _{C}\left( t\right) =\hat{c}\Psi \left( t\right) ,\ \ \Phi \left(
t,\tau \right) =\hat{U}_{t}\left( \tau \right) \Psi _{C}\left( t\right) ,
\label{notations}
\end{equation}%
We arrive at

\begin{equation}
K\left( t,\tau \right) =\left\langle \ \Phi \left( t,\tau \right) \right.
\left\vert \Psi _{C}\left( t+\tau \right) \right\rangle .  \label{K(t,tao)}
\end{equation}%
One can formulate these steps as a general procedure how to calculate the
correlator $K\left( t,\tau \right) $ from the solution to the Schr\"{o}%
dinger equation:

(a) Calculate the function $\Psi _{C}\left( t+\tau \right) =$ $\hat{c}\Psi
\left( t+\tau \right) $, where $\Psi \left( t+\tau \right) $ is the solution
to the Schr\"{o}dinger equation at the time interval $\left[ 0,t+\tau \right]
$ with initial condition $\left\vert \Psi \left( 0\right) \right\rangle $.

(b) Calculate the function $\Phi \left( t,\tau \right) $. In order to do
that, one has to solve the Schr\"{o}dinger equation at the time interval $%
\left[ t,t+\tau \right] $ with initial condition $\Psi _{C}\left( t\right) $. To calculate $\Psi _{C}\left( t\right) $, one needs 
to repeat step (a) over the time interval $\left[ 0,t\right] $
instead of $\left[ 0,t+\tau \right] $.

Note that a complete closed system including both the dynamical system and
the reservoir has some evolution operator too, say $\hat{U}_{t_{0}}\left(
t^{\prime }\right) $, even though it is unknown to us. Nevertheless, if we
arrive at Eq.~(\ref{K(t,tao)}) for the closed system, this equation would
contain both dynamical and reservoir degrees of freedom. Within the Langevin
approach which describes the system with stochastic equations of evolution,
the averaging over the reservoir degrees of freedom is equivalent to
averaging over the statistics of the noise sources \cite{Landau1965}. This
paradigm allows one to describe open systems without relying on the density
matrix.

To summarize, one can calculate the power spectrum by solving Eq.~(\ref{S eq}%
) to find the functions $\Psi _{C}\left( t \right)$, $\Psi _{C}\left( t+\tau
\right)$, and $\Phi \left( t,\tau \right) $ which depend on the noise
sources, then substitute them functions into Eq.~(\ref{K(t,tao)}) and
perform averaging over the noise statistics:
\begin{equation}
K\left( t,\tau \right) =\overline{\left\langle \ \Phi \left( t,\tau \right)
\right. \left\vert \Psi _{C}\left( t+\tau \right) \right\rangle }.
\label{K bar}
\end{equation}%
Eq.~(\ref{K bar}) follows a general scheme of calculating the observables
using the stochastic equation for the state vector: a standard procedure of
solving the Schr\"{o}dinger equation is supplemented by averaging over the
noise statistics.

%%%%%%%%%%%%%%%%%%%%%%%%%%%%%%%%%%%%%%%%%%%%%%%

\subsection{Emission spectrum of an excited qubit screened by an ensemble of
ground-state qubits}

Consider again an ensemble of $N$ identical qubits in a cavity and assume
that the cavity field decays much faster than the qubits. As before, we will
solve for the evolution over the intermediate timescales when only the field
dissipation has to be taken into account. Consider an initial state in which
only one qubit is excited, corresponding to Eq.~(\ref{Initial state}) in which one should take $C_{01} = 1$ and $C_{0j\neq 1} = 0$. As
usual, we seek the solution of the stochastic equation for the state vector
in the form of Eq.~(\ref{Psi}). From Eq.~(\ref{Psi}) and the first of Eqs.~(%
\ref{notations}) one can get%
\begin{equation}
\Psi _{C}\left( t\right) =\hat{c}\Psi \left( t\right) =C_{10}\left( t\right)
\left\vert 0\right\rangle \Pi _{j=1}^{N}\left\vert 0_{j}\right\rangle
\label{psiC}
\end{equation}

According to the above procedure, we need to find the solution of Eqs.~(\ref%
{c10 eq sp})-(\ref{C00 TD}) with initial condition (\ref{psiC}) at the time
interval $[t,t+\tau]$. One can see that Eqs.~(\ref{c10 eq sp}),(\ref{c0j eq
sp}) have a trivial zero solution, whereas Eq.~(\ref{C00 TD}) yields
\begin{equation}
\ \Phi \left( t,\tau \right) =\left[ e^{-i\frac{\omega }{2}\tau
}C_{10}\left( t\right) -\frac{i}{\hbar }e^{-i\frac{\omega }{2}\tau
}\int_{0}^{\tau }e^{i\frac{\omega }{2}t^{\prime }}\mathfrak{R}_{00}\left(
t+t^{\prime }\right) dt^{\prime }\right] \left\vert 0\right\rangle \Pi
_{j=1}^{N}\left\vert 0_{j}\right\rangle  \label{Phi}
\end{equation}%
Substituting Eqs.~(\ref{psiC}) and (\ref{Phi}) into Eq.~(\ref{K bar}) and
taking into account that the term linear with respect to the noise source
gives zero upon averaging, we obtain%
\begin{equation}
K\left( t,\tau \right) =e^{i\frac{\omega }{2}\tau }C_{10}^{\ast }\left(
t\right) C_{10}\left( t+\tau \right)  \label{k ave}
\end{equation}%
Using Eq.~(\ref{c10}) for the function $C_{10}\left( t\right)$ we get%
\begin{equation}
K\left( t,\tau \right) =\frac{\left\vert \Omega _{R}\right\vert ^{2}}{\Xi
^{2}}e^{-\left( i\omega +\frac{\mu }{4}\right) \tau }e^{-\frac{\mu }{2}%
t}\sin \left( \Xi t\right) \sin \left[ \Xi \left( t+\tau \right) \right] .
\label{k(tao,t)}
\end{equation}%
The resulting power spectrum in Eq.~(\ref{S(v)}) is given by%
\begin{equation*}
S\left( \upsilon \right) =\frac{8\left\vert \Omega _{R}\right\vert ^{2}}{\pi
\mu \left( \mu ^{2}+16\Xi ^{2}\right) }\mathrm{Re}\frac{\mu -2i\left( \nu
-\omega \right) }{\left( \frac{\mu }{4}-i\left( \nu -\omega \right) \right)
^{2}+\Xi ^{2}},
\end{equation*}%
which can be transformed using the second of Eqs.~(\ref{Solutions}) as%
\begin{equation}
S\left( \upsilon \right) =\frac{1}{2\pi }\frac{\left\vert \Omega
_{R}\right\vert ^{2}}{\left( \left( \nu -\omega \right) ^{2}-\left(
N\left\vert \Omega _{R}\right\vert ^{2}-\frac{\mu ^{2}}{8}\right) \right)
^{2}+\frac{\mu ^{2}}{4}\left( N\left\vert \Omega _{R}\right\vert ^{2}-\frac{%
\mu ^{2}}{16}\right) } .  \label{power spectrum}
\end{equation}%
Under the condition $\mu \ll 2\left\vert \Omega _{R}\right\vert \sqrt{N}$ we
can obtain%
\begin{equation*}
S( \nu) =\frac{1}{2\pi }\frac{\left\vert \Omega _{R}\right\vert ^{2}}{\left(
\left( \nu -\omega \right) ^{2}-N\left\vert \Omega _{R}\right\vert
^{2}\right) ^{2}+\frac{\mu ^{2}}{4}N\left\vert \Omega _{R}\right\vert ^{2}},
\end{equation*}%
i.e., the spectrum consists of two well-resolved lines shifted with respect
to $\omega $ by $\pm \left\vert \Omega _{R}\right\vert \sqrt{N}$, with the
maximum value $S_{\max }\left( \pm \left\vert \Omega _{R}\right\vert \sqrt{N}%
\right) =\frac{1}{\pi }\frac{2}{N\mu ^{2}}$ and linewidth $\sim \frac{\mu }{2%
}$. The dependence $S_{\max }\propto \frac{1}{N}$ reflects the destructive interference 
effect described in Section C: the probability of the photon emission by a
qubit scales as $P_{rad}\approx \frac{1}{N}$.

%%%%%%%%%%%%%%%%%%%%%%%%%%%%%%%%%%%%%%%%

\begin{figure}[htb]
\includegraphics[width=0.5\textwidth]{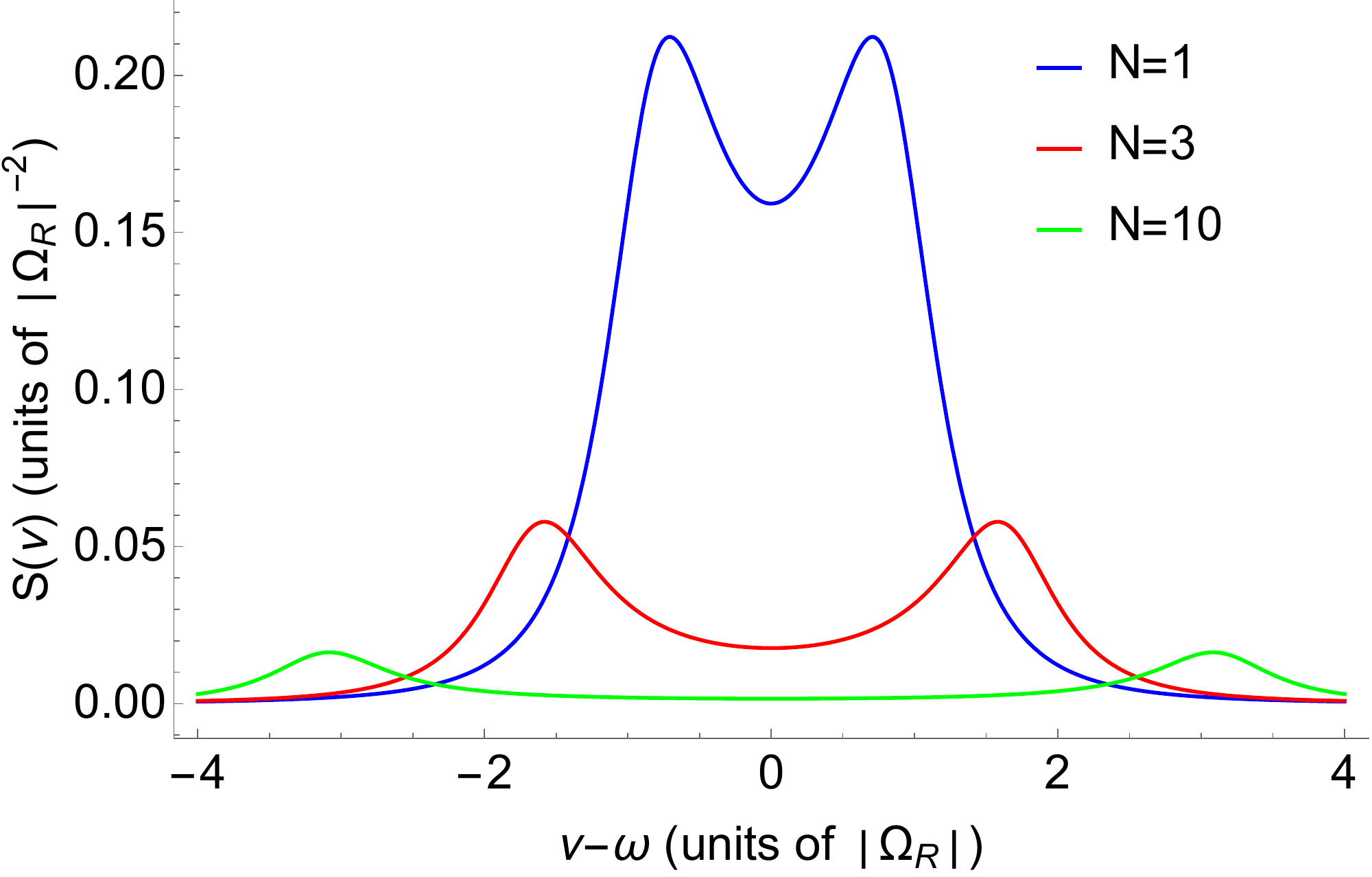}
\caption{Normalized emission spectra given by Eq.~(\protect\ref{power
spectrum}) for three values of $N$ and the cavity decay rate $\protect\mu/2
= \Omega_R$. The height of the peaks scales as $1/N$. }
\label{spectra}
\end{figure}

This behavior is illustrated in Fig.~\ref{spectra} which shows the emission
spectra given by Eq.~(\ref{power spectrum}) for three different qubit
numbers $N$ and the cavity decay rate $\mu/2 = \Omega_R$.

%%%%%%%%%%%%%%%%%%%%%%%%%%%%%%%%%%%%%%%%

\section{Many-qubit systems with different transition frequencies}

In this section we consider an ensemble of qubits with a large spread of
transition frequencies interacting with a cavity mode. This is usually the case for quantum dots where the inhomogeneous broadening is related to dispersion of their sizes. Assuming that the
inhomogeneous broadening dominates, we will neglect the homogeneous
broadening of the qubit-field resonance. We will still consider
single-photon excitations when the state vector can be represented in the
form of Eq.~(\ref{Psi}) and the coefficients satisfy Eqs.~(\ref{c10 eq sp})-(%
\ref{C00 TD}). If the condition $\left\vert \Delta _{j}\right\vert \gg \frac{%
\gamma _{j}}{2}$, $\frac{\mu }{2}$ is satisfied for most qubits, we can
neglect the terms $\frac{\gamma _{j}}{2}C_{0j}$ and $\frac{\mu }{2}C_{10}$
in Eqs.~(\ref{c10 eq sp}), (\ref{c0j eq sp}) which lead to homogeneous
broadening. To preserve the norm of the state vector, we also need to drop
the noise term in the right-hand side of Eq.~(\ref{C00 TD}). The resulting
set of equations becomes%
\begin{equation}
\dot{C}_{10}+i\frac{3\omega }{2}C_{10}-i\sum_{j=1}^{N}\Omega _{Rj}^{\ast
}C_{0j}=0,  \label{c10 eq}
\end{equation}%
\begin{equation}
\dot{C}_{0j}+i\left( \frac{3\omega }{2}+i\Delta _{j}\right) C_{0j}-i\Omega
_{Rj}C_{10}=0,  \label{c0j eq}
\end{equation}%
\begin{equation}
\dot{C}_{00}+i\frac{\omega }{2}C_{00}=0.  \label{c00 eq}
\end{equation}%
Eigenfrequencies of the system (\ref{c10 eq})-(\ref{c0j eq}) are determined in the same way as in previous
sections,
\begin{equation}
\left(
\begin{array}{c}
C_{10} \\
C_{01} \\
C_{02} \\
\vdots%
\end{array}%
\right) \propto e^{\left( -i\frac{3\omega }{2}+\Gamma \right) t},
\label{Eigenfrequencies}
\end{equation}%
where%
\begin{equation*}
\ \ \det \ \left(
\begin{array}{cccc}
\Gamma & -i\Omega _{R1}^{\ast } & -i\Omega _{R2}^{\ast } & \cdots \\
-i\Omega _{R1} & \left( \Gamma +i\Delta _{1}\right) & 0 & \cdots \\
-i\Omega _{R2} & 0 & \left( \Gamma +i\Delta _{2}\right) & \cdots \\
\vdots & \vdots & \ddots & \cdots%
\end{array}%
\right) =0,\ \ \ \
\end{equation*}%
\begin{equation}
\Gamma \Pi _{j=1}^{N}\left( \Gamma +i\Delta _{j}\right)
+\sum_{j=1}^{N}\left\vert \Omega _{Rj}\right\vert ^{2}\Pi _{m\neq
j}^{N}\left( \Gamma +i\Delta _{m}\right) =0  \label{ceq}
\end{equation}%
These equations can be solved numerically. However, some analytic progress
is possible if the spectrum of qubit frequencies is continuous.

\subsection{ Continuous spectrum}

The formal solution of Eq.~(\ref{c0j eq}) is%
\begin{equation}
G_{0j}=G_{0j}\left( 0\right) e^{-i\Delta _{j}t}+i\Omega _{Rj}e^{-i\Delta
_{j}t}\int_{0}^{t}e^{i\Delta _{j}\tau }G_{10}\left( \tau \right) d\tau .
\label{g0j}
\end{equation}%
Substituting it into Eq.~(\ref{c10 eq}) gives%
\begin{equation}
\dot{G}_{10}=i\sum_{j=1}^{N}\Omega _{Rj}^*G_{0j}\left( 0\right) e^{-i\Delta
_{j}t}-\int_{0}^{t}\sum_{j=1}^{N}|\Omega _{Rj}|^{2}e^{i\Delta _{j}\left( \tau
-t\right) }G_{10}\left( \tau \right) d\tau .  \label{g10 dot solution}
\end{equation}%
Now we can go from summation to integration by introducing
\begin{equation*}
\sum_{j=1}^{N}\left( \cdots \right) \Rightarrow \int_{-\infty }^{\infty
}\left( \cdots \right) g\left( \Delta \right) d\Delta .
\end{equation*}%
where $g\left( \Delta \right) $ is the spectral density of qubit states
which can be presented in a normalized form as%
\begin{equation}
g=\frac{N}{2\delta }f\left( \frac{\Delta }{\delta }\right) ,  \label{g}
\end{equation}%
where $2\delta $ is the width of the spectrum; $f\left( \frac{\Delta }{%
\delta }\right) \rightarrow 0$ when $\left\vert \Delta \right\vert \gg
\delta $ and $\frac{1}{2\delta }\int_{-\infty }^{\infty }f\left( \frac{%
\Delta }{\delta }\right) d\Delta =1.$

Taking the initial state as $\left\vert 1\right\rangle \Pi
_{j=1}^{N}\left\vert 0_{j}\right\rangle $ (i.e. one photon is present
whereas all qubits are in the ground state) we obtain%
\begin{equation}
\dot{G}_{10}=-\int_{0}^{t}\int_{-\infty }^{\infty }e^{i\Delta \left( \tau
-t\right) }D\left( \Delta \right) G_{10}\left( \tau \right) d\Delta d\tau ,
\label{G10 dot solution}
\end{equation}%
where%
\begin{equation*}
D\left( \Delta \right) =g\left( \Delta \right) |\Omega _{R}|^{2}\left( \Delta
\right) .
\end{equation*}%
From Eq.~(\ref{G10 dot solution}) one can obtain%
\begin{equation}
\dot{G}_{10}=-\int_{0}^{t}\tilde{D}\left( \tau -t\right) G_{10}\left( \tau
\right) d\tau ,  \label{g10 dot sol}
\end{equation}%
where%
\begin{equation}
\tilde{D}\left( \tau \right) =\int_{-\infty }^{\infty }e^{i\Delta \tau
}D\left( \Delta \right) d\Delta  \label{D tilde}
\end{equation}%
is a Fourier conjugate to the spectrum $D\left( \Delta \right) $. If we
introduce the average value of the Rabi frequency as $\left\langle |\Omega
_{R}|^{2}\right\rangle $, then using Eq.~(\ref{g}) one obtains

\begin{equation}
D\left( \Delta \right) =\frac{N\left\langle |\Omega _{R}|^{2}\right\rangle }{%
2\delta }f\left( \frac{\Delta }{\delta }\right) .  \label{D}
\end{equation}

Note that the relationship between the characteristic width $2\delta $ of
the spectrum $D\left( \Delta \right) $ and the characteristic temporal width
$\Im $ of the Fourier conjugate $\tilde{D}\left( \tau \right) $ is given by
a standard formula $\Im \times 2\delta \approx 2\pi $. Next, we derive
approximate solutions for the complex amplitudes in various limits.

Let's assume that the function $G_{10}\left( t\right) $ varies over a
certain timescale $T$, which is unknown. If the spectral width $\delta$ is
large enough,%
\begin{equation}
T\delta \gg \pi ,\ \ \ t\delta \gg \pi  \label{cd}
\end{equation}
(which implies $\Im \ll T,t$), then one can obtain from Eq.~(\ref{g10 dot
sol})
\begin{equation*}
\dot{G}_{10}\approx -G_{10}\left( t\right) \int_{-\infty }^{0}\tilde{D}%
\left( \tau \right) d\tau ,
\end{equation*}
or
\begin{equation}
\dot{G}_{10}=-\kappa G_{10},  \label{G10 dot eq}
\end{equation}
where the absorption rate is%
\begin{equation}
\kappa =\int_{-\infty }^{0}\tilde{D}\left( \tau \right) d\tau  \label{kappa}
\end{equation}

Consider now the case when we still have $t\delta \gg \pi $, but $T\delta
\ll \pi $ (i.e. $t\gg \Im \gg T$). The condition $t\gg \Im $ allows one to
transform Eq.~(\ref{g10 dot sol}) as
\begin{equation}
\dot{G}_{10}\approx -\int_{-\infty }^{t}\tilde{D}\left( \tau -t\right)
G_{10}\left( \tau \right) d\tau .  \label{g10 equ}
\end{equation}%
We will seek the solution to Eq.~(\ref{g10 equ}) in the form $%
G_{10}=ge^{\Gamma t}$, and, since $\Im \gg T\approx \Gamma ^{-1}$, we write
the right-hand side of Eq.~(\ref{g10 equ}) as a series in powers of $\Gamma
^{-1}$ by sequential integration by parts:%
\begin{equation}
\int_{-\infty }^{t}\tilde{D}\left( \tau -t\right) G_{10}\left( \tau \right)
d\tau =\left( \frac{\tilde{D}\left( 0\right) }{\Gamma }+\sum_{n=1}^{\infty
}\left( -1\right) ^{n}\left( \frac{d^{n}\tilde{D}\left( \tau \right) }{d\tau
^{n}}\right) _{\tau =0}\frac{1}{\Gamma ^{n+1}}\right) g.
\label{eq of D tilde}
\end{equation}%
From Eqs.~(\ref{g10 equ},\ref{eq of D tilde}) one can obtain the equation
for $\Gamma $:%
\begin{equation}
\Gamma ^{2}+\tilde{D}\left( 0\right) =\sum_{n=1}^{\infty }\left( -1\right)
^{n}\left( \frac{d^{n}\tilde{D}\left( \tau \right) }{d\tau ^{n}}\right)
_{\tau =0}\frac{1}{\Gamma ^{n}}.  \label{equ for gamma}
\end{equation}%
Taking into account that $\frac{d}{d\tau }\sim \frac{1}{\Im }$ , the
right-hand side of Eqs.~(\ref{equ for gamma}) is an expansion in powers of a
small parameter $\left( \Im \Gamma \right) ^{-1}$. Note that $\tilde{D}%
\left( \tau \right) $ is a Fourier conjugate for the \textit{real} spectrum $%
D\left( \Delta \right) $, therefore $\tilde{D}\left( 0\right) =\int_{-\infty
}^{\infty }D\left( \Delta \right) d\Delta $ is real. In zeroth order with
respect to $\left( \Im \Gamma \right) ^{-1}$ we obtain $\Gamma =\Gamma
_{0}=\pm i\sqrt{\tilde{D}\left( 0\right) }$. Next, we substitute $\Gamma
=\Gamma _{0}+\eta $ into Eq.~(\ref{equ for gamma}), where $\left\vert \Gamma
_{0}\right\vert \gg \left\vert \eta \right\vert $:%
\begin{equation}
\eta =\frac{\sum_{n=1}^{\infty }\left( -1\right) ^{n}\left( \frac{d^{n}%
\tilde{D}\left( \tau \right) }{d\tau ^{n}}\right) _{\tau =0}\frac{1}{\Gamma
_{0}^{n}}}{2\Gamma _{0}+\frac{1}{\Gamma _{0}}\sum_{n=1}^{\infty }\left(
-1\right) ^{n}n\left( \frac{d^{n}\tilde{D}\left( \tau \right) }{d\tau ^{n}}%
\right) _{\tau =0}\frac{1}{\Gamma _{0}^{n}}}  \label{eta}
\end{equation}%
Taking into account that $\left( \frac{d^{n}\tilde{D}\left( \tau \right) }{%
d\tau ^{n}}\right) _{\tau =0}=\int_{-\infty }^{\infty }\left( i\Delta
\right) ^{n}D\left( \Delta \right) d\Delta $, and $D\left( \Delta \right) $
is a real quantity, it is easy to see that in the numerator of Eq.~(\ref{eta}%
) \textit{all} terms are real, and in the denominator all terms are
imaginary. Therefore, $\eta $ is imaginary, i.e. it is a frequency shift,
not the decay constant. This is true in any order with respect to $\left(
\Im \Gamma \right) ^{-1}$ . In particular, consider the largest term,
\begin{equation}
\eta \approx \frac{1}{2\tilde{D}\left( 0\right) }\left( \frac{d\tilde{D}%
\left( \tau \right) }{d\tau }\right) _{\tau =0}=i\frac{\int_{-\infty
}^{\infty }\Delta D\left( \Delta \right) d\Delta }{2\int_{-\infty }^{\infty
}D\left( \Delta \right) d\Delta }  \label{Eta}
\end{equation}%
This term is of the order of $\Im ^{-1}\sim \delta $, i.e. of the order of
the spectral bandwidth.

If we use Eq.~(\ref{D}), in which the function $f\left( \frac{\Delta }{%
\delta }\right) $ has a rectangular shape, we get
\begin{equation}
\kappa \approx \pi \frac{N}{2\delta }\left\langle |\Omega
_{R}|^{2}\right\rangle ,\ \ \ \left\vert \Gamma _{0}\right\vert \approx \sqrt{%
N\left\langle |\Omega _{R}|^{2}\right\rangle }.\   \label{kap and xi}
\end{equation}

It is easy to see that the quantity $2\kappa $ corresponds \textit{exactly}
to the probability of the transition per unit time (from Fermi's Golden
Rule) from the state $\left\vert 1\right\rangle \Pi _{j=1}^{N}\left\vert
0_{j}\right\rangle $ into the continuous spectrum of states of excited
qubits. For the field-matter coupling regime described by the absorption
rate Eq.~(\ref{kappa}) we have $T\sim \kappa ^{-1}$ ; in this case the
required condition $T\delta \gg \pi $ corresponds to $\sqrt{N\left\langle
|\Omega _{R}|^{2}\right\rangle }\ll \frac{\delta }{\pi }$.

In the light-matter interaction regime corresponding to the frequency shift $%
\sim \left\vert \Gamma _{0}\right\vert $ , we have $T\sim \left\vert \Gamma
_{0}\right\vert ^{-1}$; the required condition $\ T\delta \ll \pi $ for this
regime corresponds to $\sqrt{N\left\langle |\Omega _{R}|^{2}\right\rangle }\gg
\frac{\delta }{\pi }$.

To summarize this part, the effect of the inhomogeneous broadening depends
on the relationship between the collective Rabi frequency $\sqrt{%
N\left\langle |\Omega _{R}|^{2}\right\rangle }$ and the spread of transition
frequencies $\Delta _{j}=\frac{W_{j}}{\hbar }-\omega $. If this spread is
greater than the collective Rabi frequency, the light-qubit coupling leads
to the photon absorption accompanied by the transition into the continuum of
excited-qubit states described by a standard perturbation theory. In the
opposite case, when the collective Rabi frequency exceeds the spread of
transition frequencies, the dynamics of individual qubits becomes 
synchronized and the photon interacts with an ensemble of qubits as with a
single qubit with a giant dipole moment enhanced by the factor $\sqrt{N}$.

%%%%%%%%%%%%%%%%%%%%%%%%%%%%%%%%%%%%%%%%%%

\subsection{Electron states in quantum wells}

Here we consider a multilevel quantum-confined
electron system such as electron states in a quantum well or perhaps in a quantum wire or a multilevel quantum dot.  In this case optical
transitions occur generally between two groups of electron energy states,
for example between electron states in the conduction band and valence band.
Let's take zero energy between these two groups and denote positive energies
in the conduction band as $W_{j}$ (Latin indices) and negative energies in
the valence band as $-W_{\alpha }$ (Greek indices). The frequencies of the
optical transitions are%
\begin{equation}
\omega _{j\alpha }=\frac{W_{j}+W_{\alpha }}{\hbar }.  \label{fot}
\end{equation}%
We won't consider here the intraband optical transitions within each group,
e.g. $\alpha \iff \beta $ or $m\iff n$, although the formalism below can be
easily extended to include them.

The RWA Hamiltonian is
\begin{equation}
\hat{H}=\hbar \omega \left( \hat{c}^{\dagger }\hat{c}+\frac{1}{2}\right)
+\sum_{j=1}^{J}W_{j}\hat{a}_{j}^{\dagger }\hat{a}_{j}-\sum_{\alpha
=1}^{A}W_{\alpha }\hat{a}_{\alpha }^{\dagger }\hat{a}_{\alpha }-\hbar
\sum_{j=1}^{J}\sum_{\alpha =1}^{A}\left[ \left( \Omega _{R;j\alpha }\hat{a} 
_{j}^{\dagger }\hat{a}_{\alpha }\hat{c}+\Omega _{R;j\alpha }^*\hat{a}_{\alpha
}^{\dagger }\hat{a}_{j}\hat{c}^{\dagger }\right) \right] ,
\label{RWA hamiltonian}
\end{equation}%
where $\Omega _{R;j\alpha }=\frac{\mathbf{d}_{\alpha j}\mathbf{\cdot E}}{%
\hbar }.$

Instead of the excitation and deexcitation operators for a qubit that are
specific to a two-level system, $\hat{\sigma}^{\dagger }$and $\hat{\sigma}$,
it is easier to introduce standard creation and annihilation operators of
the fermion states. Therefore, the states that were denoted as $\left\vert
0_{j\alpha }\right\rangle $ and $\left\vert 1_{j\alpha }\right\rangle $ when
using the operators $\hat{\sigma}^{\dagger }$ and $\hat{\sigma}$ become $%
\left\vert 0_{j}\right\rangle \left\vert 1_{\alpha }\right\rangle $ and $%
\left\vert 1_{j}\right\rangle \left\vert 0_{\alpha }\right\rangle $ when
using standard fermion operators.

We consider again lowest-energy states corresponding to zero- or
single-photon excitation, 
\begin{eqnarray}
\Psi & = & C_{00}\left\vert 0\right\rangle \Pi _{j=1}^{J}\left\vert
0_{j}\right\rangle \Pi _{\alpha =1}^{A}\left\vert 1_{\alpha }\right\rangle
+C_{10}\left\vert 1\right\rangle \Pi _{j=1}^{J}\left\vert 0_{j}\right\rangle
\Pi _{\alpha =1}^{A}\left\vert 1_{\alpha }\right\rangle  \notag \\
& + & \sum_{j,\alpha }^{N,A}C_{0j\alpha }\left\vert 0\right\rangle
\left\vert 1_{j}\right\rangle \left\vert 0_{\alpha }\right\rangle \Pi
_{m\neq j}^{J}\left\vert 0_{m}\right\rangle \Pi _{\beta \neq \alpha
}^{A}\left\vert 1_{\beta }\right\rangle.  \label{spe}
\end{eqnarray}%
The equations for excited state coefficients are
\begin{equation}
\dot{C}_{10}+i\left( \omega \frac{3}{2}-\sum_{\alpha }\frac{W_{\alpha }}{%
\hbar }\right) C_{10}-i\sum_{j,\alpha }^{N,A}\Omega _{R;j\alpha }^* C_{0j\alpha
}=0,  \label{c10 d}
\end{equation}%
\begin{equation}
\dot{C}_{0j\alpha }+i\left( \omega \frac{1}{2}+\frac{W_{j}}{\hbar }%
-\sum_{\beta \neq \alpha }\frac{W_{\beta }}{\hbar }\right) C_{0j\alpha
}-i\Omega _{R;j\alpha }C_{10}=0,  \label{c0jalpha dot}
\end{equation}%
Following a standard procedure, we seek the solution as
\begin{equation*}
\left(
\begin{array}{c}
C_{10} \\
C_{0j\alpha }%
\end{array}%
\right) =\left(
\begin{array}{c}
G_{10} \\
G_{0j\alpha }%
\end{array}%
\right) e^{-i\left( \omega \frac{3}{2}-\sum_{\alpha }\frac{W_{\alpha }}{%
\hbar }\right) t},
\end{equation*}
which gives
\begin{equation}
\dot{G}_{10}-i\sum_{j,\alpha }^{N,A}\Omega _{R;j\alpha }^* G_{0j\alpha }=0,
\label{g10 d}
\end{equation}%
\begin{equation}
\dot{G}_{0j\alpha }+i\Delta _{j\alpha }G_{0j\alpha }-i\Omega _{R;j\alpha
}G_{10}=0,  \label{g0jalpha dot}
\end{equation}%
where $\Delta _{j\alpha }=\omega _{j\alpha }-\omega .$

Their formal solution can be written as
\begin{equation}
G_{0j\alpha }=G_{0j\alpha }\left( 0\right) e^{-i\Delta _{j\alpha }t}+i\Omega
_{R;j\alpha }e^{-i\Delta _{j\alpha }t}\int_{0}^{t}e^{i\Delta _{j\alpha }\tau
}G_{10}\left( \tau \right) d\tau ,  \label{Gojalpha}
\end{equation}%
\begin{equation}
\dot{G}_{10}=i\sum_{j,\alpha }^{N,A}\Omega _{R;j\alpha }^* G_{0j}\left(
0\right) e^{-i\Delta _{j\alpha }t}-\int_{0}^{t}\sum_{j,\alpha }^{N,A}|\Omega
_{R;j\alpha }|^{2}G_{10}\left( \tau \right) ^{i\Delta _{j\alpha }\left( \tau
-t\right) }d\tau.  \label{G10 d}
\end{equation}

We assume that zero detuning $\Delta _{j\alpha }=0$ corresponds to some
particular values of energies $W_{j0}$ and $W_{\alpha 0}$, and introduce the
energy detunings with respect to those values:
\begin{equation*}
\Delta _{j}=\frac{W_{j0}-W_{j}}{\hbar },\ \ \ \ \Delta _{\alpha }=\frac{%
W_{\alpha 0}-W_{\alpha }}{\hbar }.
\end{equation*}%
In the limit of a continuous spectrum we obtain%
\begin{eqnarray}
 & \dot{G}_{10} = i\int \int \Omega _{R}^*\left( \Delta _{J}+\Delta _{A}\right)
g_{J}\left( \Delta _{J}\right) g_{A}\left( \Delta _{A}\right) G_{0JA}\left(
0\right) e^{-i\left( \Delta _{J}+\Delta _{A}\right) t}d\Delta _{J}d\Delta
_{A} & \notag \\
&-\int_{0}^{t}\int \int |\Omega _{R}|^{2}\left( \Delta _{J}+\Delta
_{A}\right) g_{J}\left( \Delta _{J}\right) g_{A}\left( \Delta _{A}\right)
G_{10}\left( \tau \right) e^{i\left( \Delta _{J}+\Delta _{A}\right) \left(
\tau -t\right) }d\Delta _{J}d\Delta _{A}d\tau,  \label{G10 td} &
\end{eqnarray}%
where $\sum_{j,\alpha }^{N,A}\left( \cdots \right) \Rightarrow \int \int
\left( \cdots \right) g_{J}\left( \Delta _{J}\right) g_{A}\left( \Delta
_{A}\right) d\Delta _{J}d\Delta _{A}$, $\Delta _{j\alpha }\Rightarrow \Delta
_{J}+\Delta _{A}$; the functions $g_{J}\left( \Delta _{J}\right) ,$ $%
g_{A}\left( \Delta _{A}\right) $ are corresponding spectral densities of
states.

The subsequent arguments follow those in the previous section. For example,
if the initial state has $G_{0JA}\left( 0\right) =0$, i.e., the electron
states are not excited,
\begin{equation}
\dot{G}_{10}=-\int_{0}^{t}\int \int e^{i\Delta _{J}\left( \tau -t\right)
}e^{i\Delta _{A}\left( \tau -t\right) }D\left( \Delta _{J},\Delta
_{A}\right) G_{10}\left( \tau \right) d\Delta _{J}d\Delta _{A}d\tau
\label{g10 td}
\end{equation}%
where%
\begin{equation*}
D\left( \Delta _{J},\Delta _{A}\right) =|\Omega _{R}|^{2}\left( \Delta
_{J}+\Delta _{A}\right) g_{J}\left( \Delta _{J}\right) g_{A}\left( \Delta
_{A}\right) .
\end{equation*}%
From Eq.~(\ref{g10 td}) one obtains%
\begin{equation}
\dot{G}_{10}=-\int_{0}^{t}\tilde{D}\left( \tau -t\right) G_{10}\left( \tau
\right) d\tau ,  \label{g10 time eq}
\end{equation}%
where%
\begin{equation*}
\tilde{D}\left( \tau \right) =\tilde{D}\left( \tau _{1,}\tau _{2}\right)
_{\tau _{1}=\tau _{2}=\tau }.
\end{equation*}%
\begin{equation}
\tilde{D}\left( \tau _{1,}\tau _{2}\right) =\int \int e^{i\Delta _{J}\tau
_{1}}e^{i\Delta _{A}\tau _{2}}D\left( \Delta _{J},\Delta _{A}\right) d\Delta
_{J}d\Delta _{A}  \label{d 12}
\end{equation}%
is the Fourier conjugate corresponding to the 2D spectrum $D\left( \Delta
_{J},\Delta _{A}\right) $. Introducing normalized densities of states,%
\begin{equation}
g_{J}\left( \Delta _{J}\right) =\frac{J}{2\delta _{J}}f_{J}\left( \frac{%
\Delta _{J}}{\delta _{J}}\right) ,\ \ \ g_{A}\left( \Delta _{A}\right) =%
\frac{A}{2\delta _{A}}f_{A}\left( \frac{\Delta _{A}}{\delta _{A}}\right)
\label{gj and ga}
\end{equation}%
where $2\delta _{J,A}$ are characteristic spectral widths, i.e.,
\begin{equation*}
f_{J,A}\left( \frac{\Delta _{J,A}}{\delta _{J,A}}\right) \rightarrow 0\
\mathrm{for}\, \left\vert \Delta _{J,A}\right\vert \gg \delta _{J,A}\
\mathrm{and}\ \frac{1}{2\delta _{J,A}}\int_{-\infty }^{\infty }f_{J,A}\left(
\frac{\Delta _{J,A}}{\delta _{J,A}}\right) d\Delta _{J,A}=1,
\end{equation*}%
and introducing some average value of $\left\langle |\Omega
_{R}|^{2}\right\rangle $, we obtain the 2D spectrum as
\begin{equation}
D\left( \Delta _{J},\Delta _{A}\right) =\frac{JA}{4\delta _{J}\delta _{A}}%
\left\langle |\Omega _{R}|^{2}\right\rangle f_{J}\left( \frac{\Delta _{J}}{%
\delta _{J}}\right) f_{A}\left( \frac{\Delta _{A}}{\delta _{A}}\right).
\label{DJA}
\end{equation}%
The average value of the Rabi frequency $\left\langle |\Omega
_{R}|^{2}\right\rangle $ can be estimated by averaging over the matrix
elements of the dipole-allowed optical transitions $\alpha \rightarrow j$.
The \textquotedblleft width\textquotedblright\ of the 2D spectrum $D\left(
\Delta _{J},\Delta _{A}\right) $ is of the order of $\ 2\delta _{J}\times
2\delta _{A}$. The corresponding temporal \textquotedblleft
width\textquotedblright\ of the function $\tilde{D}\left( \tau \right) =%
\tilde{D}\left( \tau _{1,}\tau _{2}\right) _{\tau _{1}=\tau _{2}=\tau }$ is
of the order of $\pi \delta _{\max }^{-1}$, where $\delta _{\max }=\max %
\left[ \delta _{J},\delta _{A}\right] .$

In what follows we repeat the same steps as in previous section, because
they are based on the same equation (compare Eqs.~(\ref{g10 dot sol}) with
Eq.~(\ref{g10 time eq})).

Consider the function $G_{10}\left( t\right) $ which varies over a certain
unknown timescale $T$. In the limit of a long $T$, namely%
\begin{equation}
T\delta _{\max }\gg \pi ,\ \ \ t\delta _{\max }\gg \pi ,  \label{limit}
\end{equation}%
Eq.~(\ref{g10 time eq}) becomes an equation for an exponentially decaying
function,%
\begin{equation}
\dot{G}_{10}=-\kappa G_{10},  \label{g10 time equation}
\end{equation}%
where the decay rate is given by%
\begin{equation}
\kappa =\int_{-\infty }^{0}\tilde{D}\left( \tau \right) d\tau  \label{dr}
\end{equation}%
As mentioned in the previous section, Eqs.~(\ref{g10 time equation}), (\ref%
{dr}) correspond to a standard transition into the states with continuous
spectrum, which is described by Fermi's Golden Rule.

In the opposite limit of a short $T$, namely $T\delta _{\max }\ll \pi $,
similarly to subsection A, we obtain the regime of a giant qubit:
\begin{equation}
G_{10}\propto e^{\Gamma t},\ \ \Gamma \approx \pm i\left\vert \Gamma
_{0}\right\vert +o\left( \frac{\delta _{\max }}{\left\vert \Gamma
_{0}\right\vert }\right)  \label{roagq}
\end{equation}%
where $\left\vert \Gamma _{0}\right\vert =$ $\sqrt{\tilde{D}\left( 0\right) }
$and the function $\tilde{D}\left( \tau \right) $ is given by Eqs.~(\ref{g10
time eq}), (\ref{d 12}). Similarly to subsection A, all subsequent terms in
the expansion of this solution in powers of the small parameter $\frac{%
\delta _{\max }}{\left\vert \Gamma _{0}\right\vert }$ are imaginary, i.e.
they do not lead to the decay of Rabi oscillations.

Within the model described by Eq.~(\ref{DJA}), if we further assume the
functions $f_{J,A}\left( \frac{\Delta _{J,A}}{\delta _{J,A}}\right) $ to
have a rectangular shape, we obtain the expressions~(\ref{Eta}) obtained in
the previous subsection, after substituting $N\Longrightarrow J\times A$ and
$\delta \Longrightarrow \delta _{\max }$. The same substitutions should be
made in the inequalities describing the conditions for various regimes:

\textbf{(i)} The regime of transitions to the continuum, corresponding to
the standard perturbation theory, is realized when $\sqrt{JA\left\langle
|\Omega _{R}|^{2}\right\rangle }\ll \frac{\delta _{\max }}{\pi }$ ;

\textbf{(ii)} The regime of a giant
qubit is realized when $\sqrt{JA\left\langle |\Omega _{R}|^{2}\right\rangle }%
\gg \frac{\delta _{\max }}{\pi }$ .

The spin-related degeneracy can be included by doubling the values of
$JA$. 

\section{Conclusions}

We found analytic solutions for the quantum dynamics of many-qubit systems strongly coupled to a quantized electromagnetic cavity mode, in the presence of decoherence and dissipation for both fermions and cavity photons.  The analytic solutions are derived for a broad class of open quantum systems including identical qubits, an ensemble of qubits with a broad distribution of transition frequencies, and multi-level electron systems. The formalism is based on the stochastic equation  of evolution for the state vector, within Markov approximation for the relaxation processes and rotating wave approximation with respect to the optical transition frequencies. This approach brings significant advantages as compared to the existing methods of treating open quantum systems. Indeed, in the master equation formalism the number of equations for $M$ degrees of freedom grows as $\frac{1}{2} M(M+1)$. Therefore, with growing number of the degrees of freedom it quickly becomes intractable analytically and has to be solved numerically.  In contrast, within our approach one has to solve a drastically smaller number of linear equations as explained in Appendix A below. This happens not only because the number of coefficients in the state vector is much smaller than the number of independent density matrix elements, but also because equations for the state vector coefficients are separable into low-dimensional uncoupled blocks whereas equations for the density matrix elements are not: they are all coupled by cross-terms. The latter is true also for the matrix elements of the Heisenberg operators. Moreover, for a nonperturbative dynamics of a  strongly coupled system the Heisenberg equations are nonlinear operator-valued equations even in the simplest single-qubit case.  

Furthermore, the explicit expression for the state vector within our approach makes the determination of entanglement or lack thereof much more transparent; it allows one to compare the resulting quantum state with Bell states and other quantum states popular in quantum information applications; it shows how the observables characterizing a given state gets affected by relaxation and noise. Another well-known approach, the Wigner-Weisskopf  method allows one to obtain the rate of decay of an averaged state vector through the imaginary parts of complex eigenenergies describing the relaxation of the populations of excited states. However, it ignores the noise component of the state vector which is determined by ground-state populations (and by excited states if the reservoir temperature is high enough). It does not predict a correct equilibrium state at long times which is determined by thermal and quantum noise. It does not conserve the norm as a result.  Unlike the Wigner-Weisskopf method, our approach preserves the noise-averaged norm of the state vector, treats the effects of noise correctly, and yields a correct equilibrium state at long times. 

We demonstrate in the analytic derivation that the interaction of an ensemble of qubits with a single-mode quantum field lead to highly unusual entangled states of practical importance, with destructive or constructive interference between the qubits depending on the initial excitation. For example, if a single excited photon interacts with an ensemble of qubits initially in their ground state, their constructive interference leads to the whole ensemble behaving as one collective dipole. However, if one or a small fraction of qubits were excited initially whereas he field was in the vacuum state, the subsequent relaxation drives the whole ensemble of qubits into an entangled dark state which is completely decoupled from the leaky cavity mode. Only a small fraction $~ 1/N$ of the initial excitation energy is lost before the system goes into the dark state, where $N$ is the number of qubits in the ground state. This effect is due to destructive interference in the resulting entangled state of the qubits. 

We find the regimes in which an inhomogeneously broadened ensemble of qubits or a multi-electron system with a broad distribution of transition frequencies couple to the quantized cavity field as a giant collective dipole. We derived the criteria when this regime of a collective dipole is destroyed by the inhomogeneous broadening of their frequencies, which is important for many realistic systems. The analytic solutions describing the evolution of the system are obtained in all limiting cases. These results are generalized beyond two-level systems to electron-hole excitations in quantum wells or other semiconductor nanostructures.

%%%%%%%%%%%%%%%%%%%%%%%%%%%%%%%%%%%%%%%%%%%%

\begin{acknowledgments}
This work has been supported in part by the Air Force Office for Scientific Research 
Grant No.~FA9550-17-1-0341, National Science Foundation Award No.~1936276, and Texas A\&M University through STRP, X-grant and T3-grant programs.
 M.T.~and M.E.~acknowledge the support by the Center of Excellence
``Center of Photonics'' funded by The Ministry of
Science and Higher Education of the Russian Federation,
Contract No.~075-15-2020-906.

\end{acknowledgments}

%%%%%%%%%%%%%%%%%%%%%%%%%%%%%

\appendix

\section{Evolution of the states with arbitrary excitation energies}

For an arbitrary quantum state of the N-qubit system coupled to a cavity mode the state vector can be expanded over all possible combinations of subsystems as 
\begin{equation}
\label{a1} 
\Psi = \sum_{n=0}^{\infty} \sum_{p = 0}^N \sum_{\alpha_p = 1}^{\mathcal{C}_N^p} C_{n p \alpha_p} | n \rangle |p, \alpha_p \rangle , 
\end{equation}
where $| n \rangle$ is a Fock state of the boson (EM) field and $|p, \alpha_p \rangle$ is a fermion state. Here index $\alpha_p$ denotes {\it different} subsets of $p$ elements out of a set of $j = 1,2,\dots N$, which correspond to the excitation of $p$ qubits out of $N$. The total number of such subsets is determined by the binomial coefficient $\mathcal{C}_N^p = \frac{N!}{p! (N-p)!}$. The state $|p, \alpha_p \rangle$ can be written as 
\begin{equation}
\label{a2} 
|p, \alpha_p \rangle = \left( \prod_{j_p \in \alpha_p} \left\vert \hat{\sigma}_{j_p}^{\dagger} \right\rangle \right) |0_{qub} \rangle,
\end{equation}
where $j_p \in \alpha_p$ are qubit numbers belonging to the subset marked by index $\alpha_p$ and  
\begin{equation}
\label{a3} 
|0_{qub} \rangle = \prod_{j=1}^{N} |0_j \rangle.
\end{equation} 

As expected, the energy of the given state $|n,p,\alpha_p \rangle$ does not depend on the index $\alpha_p$:
\begin{equation}
\label{a4} 
E_{n,p,\alpha_p } = \hbar\left( \frac{1}{2} + n \right) \omega + pW.
\end{equation} 

Substituting the state vector (\ref{a1}) into the Schr\"{o}dinger equation with the Hamiltonian (\ref{RWAH}) leads to a set of linear equations for the probability amplitudes  
$C_{n p \alpha_p}$, which is split into independent blocks corresponding to the condition
\begin{equation}
\label{a5} 
n+p = m = {\rm const.}
\end{equation} 
For $m \leq N$ each block contains $\sum_{p = 0}^{m}  \mathcal{C}_N^p$ equations which have the form
\begin{eqnarray}
& \dot{C}_{n p \alpha_p} + i \left[  \left( \frac{1}{2} + n \right) \omega + p\frac{W}{\hbar} \right] C_{n p \alpha_p} - & \nonumber \\
& i \left( \sqrt{n+1} \Omega_R \sum_{\alpha_{p-1}}^p C_{(n+1)(p-1) \alpha_{p-1}} + \sqrt{n} \Omega_R^* \sum_{\alpha_{p+1}}^{N-p} C_{(n-1)(p+1) \alpha_{p+1}} \right) = 0. &
\label{a6} 
\end{eqnarray} 

For given values of $n$ and $p$ satisfying Eq.~(\ref{a5}), Eq.~(\ref{a6}) corresponds to $ \mathcal{C}_N^p$ equations for different subsets $\alpha_p$. Note that in the equation for $C_{m 0 \alpha_0}$ the index $\alpha_0$ corresponds to only one possible subset with zero energy of the fermionic subsystem:  $ \mathcal{C}_N^0 = 1$, i.e., $|0,\alpha_0\rangle = |0_{qub} \rangle$. In the sums  $\sum_a^b (\dots )$ the lower index shows the type of a subset and the upper index shows the number of elements in the sum. Equation (\ref{a5}) implies that the subsets $\alpha_{p-1}$ and $\alpha_{p+1}$ are related to the subset $\alpha_p$ through 
\begin{equation}
\nonumber
|p, \alpha_p \rangle = \hat{\sigma}_{j_(p-1)}^{\dagger} \left\vert p-1, \alpha_{p-1} \right\rangle; \quad |p, \alpha_p \rangle = \hat{\sigma}_{j_(p+1)} \left\vert p+1, \alpha_{p+1} \right\rangle,
\end{equation}
where each pair $\alpha_p,\alpha_{p-1}$ or $\alpha_p,\alpha_{p+1}$ corresponds to a certain value of the qubit index: $j_{p-1}$ or $j_{p+1}$. Each subset $\alpha_p$ corresponds to a certain finite number of subsets $\alpha_{p-1}$ or $\alpha_{p+1}$ which contribute to the summation in Eq.~(\ref{a6}). 

For $m > N$ the number of equations in each block stops growing and is equal to $\sum_{p = 0}^{N}  \mathcal{C}_N^p$.  In this case it is convenient to write down separately the equation corresponding to the state with maximum energy of the fermionic subsystem, equal to $N\times W$: 
\begin{eqnarray}
\dot{C}_{(m-N) N \alpha_N} + i \left[  \left( \frac{1}{2} + m-N \right) \omega + N\frac{W}{\hbar} \right] C_{(m-N) N \alpha_N} - \nonumber \\
i \sqrt{m-N} \Omega_R \sum_{\alpha_{N-1}}^N C_{(m-N+1)(N-1) \alpha_{N-1}}  = 0,
\label{a7} 
\end{eqnarray} 
where the index $\alpha_N$ corresponds to the only possible subset ($ \mathcal{C}_N^N = 1$) for the state $\left\vert N, \alpha_N \right\rangle = \prod_{j=1}^N |1_j \rangle$. 

Solving the resulting set of equations in order to determine all probability amplitudes and therefore all observables is much easier with this method  than with any other existing approach that we are aware of. As an example, let's consider initial conditions which limit the number $m \leq M$ where $M > N$. In this case the density matrix equation leads to a set of coupled equations for independent matrix elements; the total number of equations is equal to 
$$
\frac{1}{2} \left[ \left( \sum_{p=0}^{N} (M-p) \mathcal{C}_N^p \right) + 1 \right] \sum_{p=0}^{N} (M-p) \mathcal{C}_N^p. 
$$
This number is obviously much higher than the maximum number of equations for the state vector coefficients within each block, which is equal to $\sum_{p = 0}^{N}  \mathcal{C}_N^p$. 

This drastic reduction in the number of equations to solve is not only because the number of coefficients in the state vector is much smaller than the number of independent density matrix elements; {\it it is also because equations for the state vector coefficients are separable into uncoupled blocks whereas equations for the density matrix elements are not: they are all coupled by cross-terms. The latter is true also for the matrix elements of the Heisenberg operators. } Moreover, for a strongly coupled system the Heisenberg equations are nonlinear operator-valued equations even in the simplest single-qubit case.  

For the Hamiltonian (\ref{RWAH}) the rank of the problem (the size of each block) can be further reduced significantly. Indeed, Eq.~(\ref{a6}) can be summed over all subsets, $\sum_{\alpha_p}^{\mathcal{C}_N^p}(\dots )$. This results in the following equations for the variables $F_{np} = \sum_{\alpha_p}^{\mathcal{C}_N^p}C_{n p \alpha_p}$:
\begin{eqnarray} & 
\dot{F}_{n p} + i \left[  \left( \frac{1}{2} + n \right) \omega + p\frac{W}{\hbar} \right] F_{n p} - & \nonumber \\
& i \left( \sqrt{n+1} (N-p+1) \Omega_R F_{(n+1)(p-1) } + \sqrt{n} \Omega_R^* (p+1) F_{(n-1)(p+1)} \right) = 0, &
\label{a8} 
\end{eqnarray} 
where the indices $n$ and $p$ are related by Eq.~(\ref{a5}). Therefore, the number of independent equations within each block is equal to $m+1$ if $m \leq N$ and is equal to $N+1$ if $m > N$. Each equation contains only two variables if $n = m$ or $n = 0$, or $p = M$. If none of these conditions are satisfied, then each equation contains three variables.

By far the most popular and accessible states are single-photon excitations. They are obtained from Eq.~(\ref{a8}) by taking $ m= 1$. This gives rise to the second-order set of equations (\ref{C10 dot F}) and (\ref{F dot}) from the main text.

%%%%%%%%%%%%%%%%%%%%%%%%%%%%%%%%%

\end{document}